\documentclass[twocolumn]{aastex62}
\usepackage{placeins}
\usepackage{mathtools, nccmath}
\usepackage{siunitx,etoolbox}
\usepackage{capt-of}

\usepackage{hyperref}
\usepackage{longtable}

\usepackage{amsmath,bm}
\graphicspath{{./}{figures/}}

\shortauthors{Dumont et al.}

\begin{document}

\title{Investigating the Dark Matter Halo of NGC~5128 using a Discrete Dynamical Model}

\correspondingauthor{Antoine Dumont}
\email{antoine.dumont.neira@gmail.com}

\author[0000-0003-0234-3376]{Antoine Dumont}
\affil{Max Planck Institute for Astronomy, K{\"o}nigstuhl 17, 69117 Heidelberg, Germany}

\author[0000-0003-0248-5470]{Anil C. Seth}
\affiliation{Department of Physics and Astronomy, University of Utah\\
115 South 1400 East, Salt Lake City, UT 84112, USA}

\author[0000-0002-1468-9668]{Jay Strader}
\affiliation{Department of Physics and Astronomy Michigan State University Biomedical \& Physical Sciences\\567 Wilson Rd, Room 3275 East Lansing, MI 48824-2320}

\author[0000-0003-4102-380X]{David J. Sand}
\affil{Steward Observatory, University of Arizona, 933 North Cherry Avenue, Tucson, AZ 85721, USA}

\author[0000-0001-6215-0950]{Karina Voggel}
\affiliation{Universite de Strasbourg, CNRS, Observatoire astronomique de Strasbourg, UMR 7550, F-67000 Strasbourg, France}

\author[0000-0002-1718-0402]{Allison K. Hughes}
\affil{Steward Observatory, University of Arizona, 933 North Cherry Avenue, Tucson, AZ 85721, USA}

\author[0000-0002-1763-4128]{Denija Crnojevi\'{c}}
\affiliation{University of Tampa, 401 West Kennedy Boulevard, Tampa, FL 33606, USA}

\author[0000-0001-5590-5518]{Duncan A. Forbes}
\affiliation{Centre for Astrophysics and Supercomputing, Swinburne University of Technology, Hawthorn VIC 3122, Australia}

\author[0000-0002-3856-232X]{Mario Mateo}
\affiliation{Department of Astronomy, University of Michigan, Ann Arbor, MI 48109, USA}

\author[0000-0003-0256-5446]{Sarah Pearson}\thanks{Hubble Fellow}
\affiliation{Center for Cosmology and Particle Physics, Department of Physics, New York University, 726 Broadway, New York, NY 10003, USA}

\begin{abstract}
As the nearest accessible massive early-type galaxy, NGC~5128 presents an exceptional 
opportunity to measure dark matter halo parameters for a representative elliptical galaxy. Here we take advantage of rich new observational datasets of large-radius tracers to perform dynamical modeling of NGC~5128, using a discrete axisymmetric anisotropic Jeans approach with a total tracer population of nearly 1800 planetary nebulae, globular clusters, and dwarf satellite galaxies extending to a projected distance of $\sim250$ kpc from the galaxy center.
We find that a standard NFW halo provides an excellent fit to nearly all the data, excepting a subset of the planetary nebulae that appear to be out of virial equilibrium.
The best-fit dark matter halo has a virial mass of ${\rm M}_{vir}=4.4^{+2.4}_{-1.4}\times10^{12} {\rm M}_{\odot}$, and NGC~5128 appears to sit below the mean stellar mass--halo mass and globular cluster mass--halo mass relations, which both predict a halo virial mass closer to ${\rm M}_{vir} \sim 10^{13} {\rm M}_{\odot}$. The inferred NFW virial concentration is $c_{vir}=5.6^{+2.4}_{-1.6}$, nominally lower than $c_{vir} \sim 9$ predicted from published $c_{vir}$--${\rm M}_{vir}$ relations, but within the $\sim 30\%$ scatter found in simulations. The best-fit dark matter halo constitutes only $\sim10\%$ of the total mass at 1 effective radius but $\sim50\%$ at 5 effective radii.
The derived halo parameters are relatively insensitive to reasonable variations in the 
tracer population considered, tracer anisotropies, and system inclination. Our analysis highlights the value of comprehensive dynamical modeling of nearby galaxies, and the importance of using multiple tracers to allow cross-checks for model robustness.
\end{abstract}


\section{Introduction\label{sec:intro}} 

Large-scale observations of the Universe show that more than $80\%$ of the total mass in the Universe is in the form of dark matter \citep[e.g][]{Planck2018}. At smaller scales, galaxies' rotation curves derived from neutral hydrogen (HI) observations have also revealed massive dark matter halos in spiral galaxies \citep[e.g.][]{Rubin1978,Bosma1978,Bosma1981B,1Sofue999,Sofue2001,Martinsson13}. These observations have also shown that the dark matter halo of individual spiral galaxies correlate with their luminosity, stellar mass, and possibly galaxy size \citep[for a more detailed discussion see][]{Wechsler2018}.

However, studying the dark matter halos of elliptical galaxies presents a major challenge. Unlike spiral galaxies, which have a disk-like structure rich in of HI gas to study their rotation curve, elliptical galaxies lack such a feature. As a result, directly measuring the rotation curve of these galaxies is difficult, and instead, we rely on stellar kinematics to determine the distribution of mass in their halos. Despite the potential of stellar kinematics, the faint nature of the outer parts of the elliptical galaxy's halos, makes measuring their dynamics still a challenge \citep{Brodie2014}. Nonetheless, constraints on the inner parts of dark matter halos of early-type galaxies can be obtained from integral field spectroscopy \citep{Cappellari2013} and strong-lensing \citep{Auger2010}. 

The use of abundant, bright discrete tracers such as planetary nebulae (PNe) and globular clusters (GCs) have been identified as a powerful tools to study the outer halo of elliptical galaxies, and the galaxy's gravitational potential \citep[e.g.,][]{Zhu2016,Alabi2016}. PNe are the final evolutionary stage of some low-to-intermediate-mass stars, emitting bright narrow [O III] lines that make them well-suited for spectroscopy at cosmic distances where spectroscopy of individual main sequence or giant stars is infeasible. GCs serve as tracers of the galaxy's potential extending to even larger radii, providing valuable information about the distribution of dark matter in the outer halo.

NGC~5128, also known as the Centaurus A, is among closest and brightest elliptical galaxies to the Milky Way, located at a distance of 3.8 Mpc \citep{Harris2010}. Due to its proximity, NGC~5128 has been a principal target for the study of GCs and PNe, hosting a rich population of both \citep{Harris2002,Harris2004,martini2004,Peng2004GCs,Peng2004,Woodley2005,Woodley2007,Woodley2010,Rejkuba2007,Walsh2015,Dumont2022,Hughes2022}. The kinematics of GCs and PNe in NGC~5128 have been used to study the dynamics of the galaxy's outer halo, providing important insights into the mass distribution in the galaxy \citep{Hui1995,Peng2004,Woodley2007,Woodley2010,Pearson2022,Hughes2022}. Published estimates suggest that the total enclosed of NGC~5128 is $(2.5\pm 0.3) \times 10^{12} {\rm M}_{\odot}$ within $\sim 120$ kpc based on its GC kinematics \citep{Hughes2022}, or $(1.0\pm 0.2)   \times 10^{12} {\rm M}_{\odot}$ within $\sim 90$ kpc using the PNe as kinematic tracers \citep{Woodley2007}. Based on the modeling of the tidal stellar stream of its dwarf satellite galaxy Dw3 at a projected radius of $\sim$80~kpc , \citet{Pearson2022} suggest a lower limit for its mass halo of  ${\rm M}_{200} \geq 4.7\times10^{12}$ M$_{\odot}$.

The kinematics and radial density profiles of PNe and GCs are not expected to be the same, therefore, using them both as independent kinematic tracers can help break the mass--anisotropy and the luminous-to-dark matter mass degeneracies \citep[e.g.][]{Napolitano2014}. This is especially true in the outer halo of elliptical galaxies like NGC~5128 that show signs of a complicated merging history \citep{Israel1998,Wang2020}. 
In this work, we use the most current catalogs of available kinematic tracers in NGC~5128 (stellar kinematics, GCs, PNe and dwarf satellite galaxies ), and model them without binning using discrete dynamical models. This approach helps avoid the loss of valuable information \citep[][]{Watkins2013}.

The paper is organized as follows: in Section~\ref{sec:Kinematic_Tracers}, we discuss the kinematic tracers available to constrain our dynamical model. Section~\ref{sec:Discrete_model} provides an overview of the discrete dynamical model employed in this study, including a detailed description of our Fiducial model, its assumptions, and free parameters. Section~\ref{sec:Model_results} presents the best-fit parameters of the Fiducial model, including its dark matter halo, and an assessment of the impact of systematics. This model is compared with previous measurements in Section~\ref{sec:compare}.  Then in Section~\ref{sec:discussion}, we discuss the implications of our best-fit mass model and how the dark matter halo of NGC~5128 compares to that of other elliptical galaxies.

Throughout this paper, we assume a distance of $3.8$ Mpc to NGC~5128 ($(m-M)_{0} = 27.91$; \citealt{Harris2002}). At this distance $1' \approx 1.1$ kpc. We also assume that the photometric major axis (PA) is $35^{\circ}$ east from north based on results of \citet{Silge2005},a systemic velocity for NGC~5128 of $540$ km s$^{-1}$ based on results from \citet{Hughes2022}, and an effective radius of $R_{eff} = 305\arcsec$ adopted from \citet{Dufour1979}.  

\section{Kinematic Tracers}
\label{sec:Kinematic_Tracers}
In this section we describe the available kinematic data for NGC~5128 used to constrain our dynamical mass models.  There are a total of 1267 PNe line of sight (LOS) velocities compiled by \citet{Walsh2015}, 645 confirmed GCs with LOS velocities compiled by \citet{Hughes2022}, and LOS velocities for 28 satellite dwarf satellite galaxies  \citep{Crnojevic2019,Muller2021b}.  We describe these samples each in greater detail below.  
 
The 1267 spectroscopically confirmed PNe with radial velocities compiled by \citet{Walsh2015} also include measurements from previous works \citep{Hui1995,Peng2004}, which have larger velocity errors. For our dynamical models, we only consider the 1135 PNe with smaller errors, as described in more detail in \citet{Walsh2015}.  The median projected galactocentric radius of the PNe is 9~kpc with some objects as far out as 39~kpc.  Typical velocity errors are 1.7~km~s$^{-1}$.  The kinematics of the PNe are complex with rotation around the minor axis at small radii, but with a changing axis of rotation at large radii that suggest either non-virial motions or a triaxial potential; their  kinematics are discussed in detail in \citet{Peng2004}.  

The largest, most up-to-date catalog of GCs in NGC~5128 is presented in \citet{Hughes2022}.  This cluster catalog includes 122 new measurements and compiles measurements of 523 additional literature GC velocities from a wide range of sources \citep{Graham1980,Bergh1981,Hesser1986,Harris2002,Harris2004,Peng2004GCs,martini2004,Woodley2005,Woodley2007,Woodley2010,Rejkuba2007,Beasley2008,Georgiev2009,Taylor2010,Taylor2015,KV2020,Fahrion2020,Dumont2022}.  The newly discovered GCs from \citet{Hughes2022} extend to larger radii than previous measurements, with the full sample having a median radius of 17.8~kpc and a maximum radius of $\sim 130$ kpc. The dataset has both low and high resolution measurements, with median velocity errors of 29.5~km~s$^{-1}$.  The 523 literature objects are a smaller number than the catalog of 563 GC velocities in \citet{Woodley2010} due to the discovery that many previous GC candidates were foreground stars based on Gaia measurements and subsequent velocity measurements \citep{Hughes2021,Dumont2022, Hughes2022}.

The GC population of NGC~5128 can be divided into metal-poor GCs (referred to as ``blue") and metal-rich GCs (referred to as ``red"), which have distinct spatial distributions. For our dynamical models, we treat the blue and red GC populations as separate kinematic tracers. We adopt the same classification for red and blue GCs as used in \citet{Hughes2022}, which is based on their $(u-z)_{0}$ and $(g-z)_{0}$ color. Nine GCs with velocities lacked color information and are not used in our analysis. Additionally, \citet{Hughes2022} identifies 18 GCs associated with either dwarf satellite galaxies or with other stellar streams.  For our Fiducial model we fit red and blue GCs separately, and exclude GCs associated with stellar sub-structures as these give multiple measurements for a single dynamical tracer.  In total we fit 618 GCs, 279 red GCs and 339 blue GCs.

NGC~5128 hosts 42 confirmed dwarf satellite galaxies (hereafter referred as ``dwarfs''), as compiled in \citet{Crnojevic2019} and \citet{Muller2021b}, with 28 of them having line-of-sight (LOS) velocities. We are only interested in fitting satellite galaxies that are located within the virial radius of NGC~5128, rather than orbiting in the larger group potential. Previous 
measurements suggested a virial radius in the range 200--300 kpc \citep{Peng2004,Muller2022};  therefore, we limit our sample to the 16 satellite galaxies with projected distances of $\leq 250$ kpc. 

\begin{figure}[h!]
    \centering
	\plotone{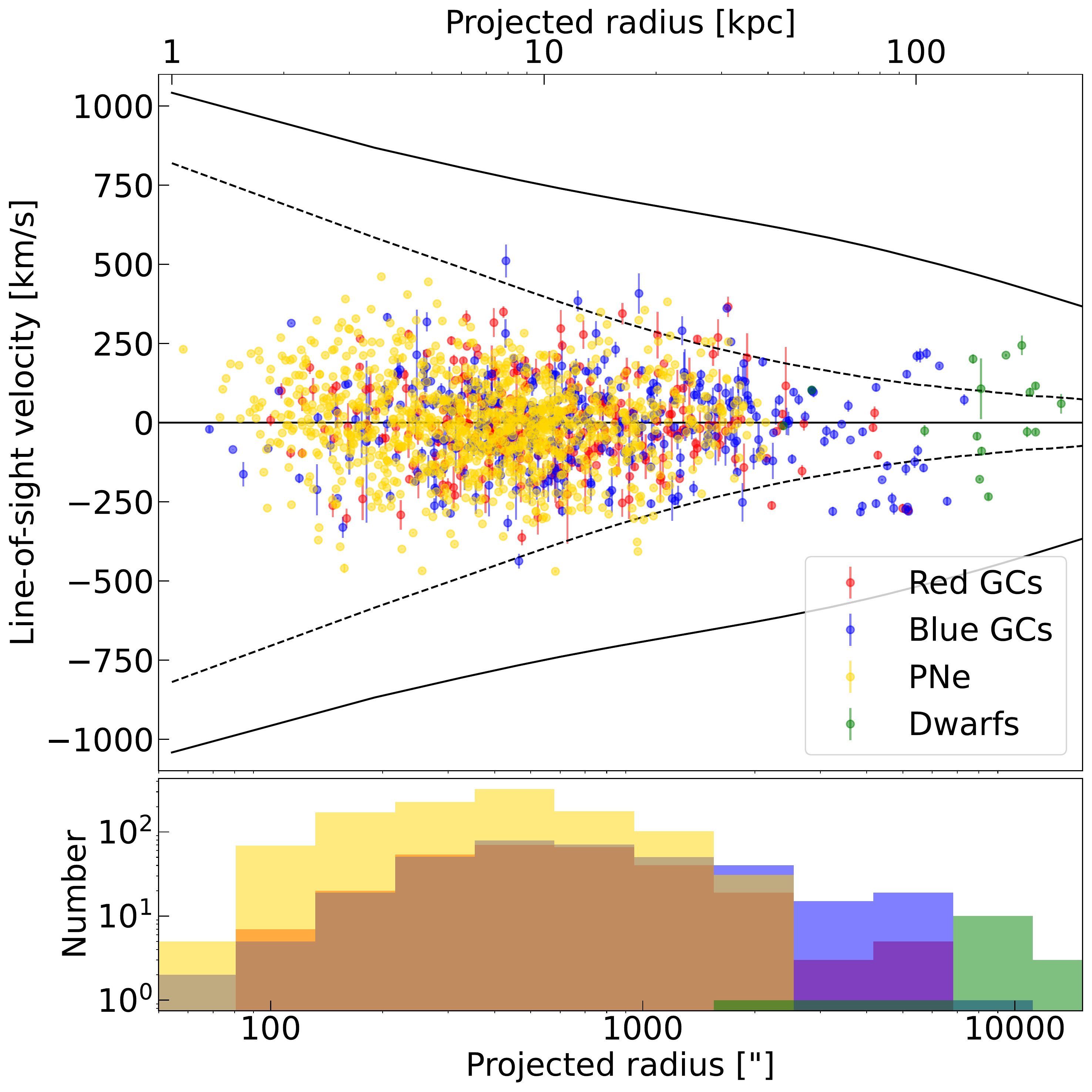}
	\caption{Radial distribution of the LOS velocity for the PNe (yellow), metal-rich and metal-poor GCs (red \& blue respectively), and the dwarf satellite galaxies (green) around NGC~5128. The LOS velocities are with respect to the systemic velocity of NGC~5128 of 540~km~s$^{-1}$ \citep{Hughes2022}. The solid black line shows the radial profile for the escape velocity for NGC~5128 derived from our best-fit mass model including dark matter presented in \S\ref{subsec:Fiducial_results}, while the dashed lines shows the escape velocity for just the stellar mass.  The bottom panel is a histogram with the radial distribution for the PNe, GCs and dwarfs in the same colors as the top panel. }
	\label{fig:Tumpet plot}
\end{figure}

The LOS velocities of our three kinematic tracers (PNe, GCs, and dwarf satellite galaxies) are shown in Figure~\ref{fig:Tumpet plot}.

Finally, we note that there is available stellar kinematic data for NGC~5128 itself from several different sources for the innermost regions of the galaxy (e.g., \citealt{Silge2005,Cappellari2009}).  Given that our primary interest is in the mass distribution at larger radii, we do not explicitly model these central data. Instead, we fix the inner ($\leq 180\arcsec$) K-band mass-to-light ratio based on the results from the stellar kinematic modeling of \citet{Cappellari2009}. A more detailed discussion on this is presented in the next section (Section \ref{subsec:grav_potential}).

\section{Discrete Dynamical Modeling}
\label{sec:Discrete_model}
In this section, we introduce the discrete dynamical modeling approach that we use to estimate NGC~5128's dynamical mass and dark matter halo properties. We will discuss the input ingredients, including models of the gravitational potential, the density profiles of each kinematic tracer, the log-likelihoods used to estimate the best-fit parameters, and finally, the Markov-Chain Monte Carlo (MCMC) method used for parameter estimation. 

We consider that the PNe, GCs (blue and red), and dwarf satellite galaxies are independent populations, each with their own spatial and dynamical properties, whose kinematics trace the same underlying gravitational potential. We model the dynamics of each population with an axisymmetric anisotropic Jeans model \citep[as implemented in the axisymmetric version of the Jeans Anisotropic Modeling software JAM;][]{JAM_cappellari}. The discrete kinematic tracers are modeled directly without binning to avoid loss of information, following the discrete maximum-likelihood modeling approach extension to the JAM software (CJAM) introduced by \citet{Watkins2013}. CJAM predicts the full three-dimensional matrix of the first and second velocity moment as a function of the major and minor axes position ($x'$ and $y'$) of the discrete kinematic tracer. Our kinematic tracers do not have available proper motions but only LOS velocity, so we only use the LOS ($z'$) component of the CJAM to constrain our models. This model assumes that the velocity ellipsoid of the kinematic tracers is aligned with the cylindrical coordinates ($x'$,$y'$,$z'$). 

\textsc{CJAM} requires the following properties as input:

\begin{itemize}
    \item The gravitational potential profile in the form of an elliptical Multi-Gaussian Expansion (MGE) \citep{MultiGaussian_Cappellari}. For our models we assume the gravitational potential is formed from the combination of luminous matter and dark matter (see more details \ref{subsec:grav_potential}). 
    \item A mass-to-light ratio ($\Upsilon$) that converts the stellar luminosity profile into a stellar mass profile. 
    \item The projected density profile for each kinematic tracer, also expressed in the form of an MGE. 
    \item The global inclination of the system $i$, which we assume to be $i=90^{\circ}$ or edge-on. While a common simplifying assumption in dynamical studies of early-type galaxies \citep[e.g][]{Zhu2016}, it is specifically supported for NGC~5128 by previous dynamical work on the PNe population \citep{Hui1995,Peng2004}. However, we do explore the impact of the assumed inclination in Section \ref{subsec:inclination}. 
    \item A rotation parameter $\kappa$ for each tracer MGE component, which we assume to be zero (no rotation) to reduce computing time (see additional discussion in Section \ref{subsec:trac}).
    \item A velocity anisotropy parameter $\beta = 1-\overline{v_z^2}/\overline{v_R^2}$ \citep[defined in cylindrical coordinates; called $\beta_z$ in][]{Cappellari2009}, which takes negative values for tangential  anisotropy and positive values for radial  anisotropy. The $\beta$ can be separately specified for each tracer MGE component.  
\end{itemize}

\subsection{Gravitational Potential}
\label{subsec:grav_potential}
We model the gravitational potential of the galaxy assuming it is made by two components; a luminous and dark matter component. We model the luminous mass potential by assuming mass follows light, characterizing its surface brightness profile (Section~\ref{subsec:lum_grav_pot}), and fitting for a mass-to-light ratio.  For the dark component (Section~\ref{subsec:dm_potential}) we assume it follows a generalized Navarro–Frenk–White (NFW) density profile \citep{Navarro1996, Zhao1996}. 

\subsubsection{Stellar Gravitational Potential}
\label{subsec:lum_grav_pot}
We characterize the contribution of luminous mass to the gravitational potential in our dynamical models by deriving the galaxy's surface brightness profile and multiplying this by a mass-to-light ratio to convert it into a mass profile. 
We derive the central surface brightness profile of NGC~5128 using data from the 2MASS Large Galaxy Atlas imaging data \citep{Jarrett2003}. 

We use the $K_s$-band mosaic for NGC~5128 since it is sensitive to the main, primarily old stellar population and is less affected by dust extinction than images at shorter wavelengths. 
The 2MASS imaging data extends to approximately $\sim 200\arcsec$ from the galaxy center. However, there is observational evidence that luminous matter contributes significantly to the gravitational potential of the galaxy inside $\sim 10$ kpc, or $\sim 600\arcsec$ \citep{Peng2004}, beyond the sensitivity limit of the 2MASS images.

To characterize the outer stellar halo, we turn to previous literature measurements by \citet{Dufour1979} and \citet{Rejkuba2022}, and combine their outer surface brightness profile data with the 2MASS K$_{s}$ image at smaller radii.  \citet{Dufour1979} derived a $V$ and $B$ surface brightness profile from unresolved stars using photographic plates out to a radius of $\sim$8\arcmin\ centered on the nucleus of NGC~5128. \citet{Dufour1979} found that the galaxy's $V$-band surface brightness profile is well represented by a de Vaucouleur's law at radial distances between $70\arcsec$ and $255\arcsec$.  Because we are deriving the surface brightness profile in the 2MASS $K_s$ band, we need to convert this profile using a $V-K_s$ color.  Based on PARSEC single stellar population models \citep{PARSEC-COLIBRI}, we expect a $V-K_s \sim 3$ for old age populations at somewhat subsolar metallicities. This color provides a good overlap between the outer part of the 2MASS image and the inner part of the \citet{Dufour1979} profile ($70\arcsec$), and we assume this color for both this profile ($\mu_{K}$ in equation~\ref{eq:dufour_SB}) and to convert the \citet{Rejkuba2022} profile below. The $V$ and $K_s$ profiles from \citet{Dufour1979} that we use is thus:
\begin{equation}
\begin{split}
    \label{eq:dufour_SB}
%      &\mu_{V}= 8.32[(a/R_{eff})^{1/4} - 1] + 22.0 \\ 
    &\mu_K = 8.32[(a/R_{eff})^{1/4} - 1] + 22.0 - 3.0
\end{split}
\end{equation}
where $R_{eff} = 305\arcsec (\sim 5.6$ kpc) is the effective radius of NGC~5128, $a= x'^{2} + y'^{2}/(1-e)^{2}$ is the major axis distance, and the the mean ellipticity of NGC~5128 is $e=1-(b'/a') = 0.23$, where $b'/a'$ is the minor-to-major axis ratio \citep{Dufour1979}.

\citet{Rejkuba2022} derive a number density profile of resolved red giant branch (RGB) stars from a large number of archival {\em Hubble} Space Telescope fields in the outer halo of NGC~5128 extending to radii of $\sim140$ kpc. They fit this data to a de Vaucouleurs-S\'ersic profile, and provide a conversion from number density to $V$-band surface brightness.  
Specifically we use equation 13 from \citet{Rejkuba2022}:

\begin{equation}
    \label{eq:Rejkuba_RGB_profile}
    \log(\Phi_{RGB}) = (7.685 \pm 0.271) - (3.008 \pm 0.142)(R/R_{eff})^{1/4}
\end{equation}
where $\Phi_{RGB}$ is the number density of RGB stars per arcmin$^{-2}$.
We then need to convert equation~\ref{eq:Rejkuba_RGB_profile} from counts per arcmin$^{-2}$ to $K_s$-band surface brightness to provide a uniform band for our input mass-model. Using the conversion given by \citet{Rejkuba2022} of  $\log(\Phi_{RGB}) = 0$ is equal to a V-band surface brightness of $\mu_{V} = 33.82$ mag arcsec$^{-2}$ and assuming $V-K_s = 3$ mag, the resulting $K_s$-band surface brightness profile in units of mag arcsec$^{-2}$ is: 
\begin{equation}
    \label{eq:Rejkuba_SB}
    \mu_{K} = -2.5\log(\Phi_{RGB}) + 30.82
\end{equation}

We assume a  PA $= 35^{\circ}$ for the two outer halo surface brightness profiles.  This is the PA found for the 2MASS image in \citet{Silge2005} and also is consistent with the PA found by \citet{Dufour1979}.  We use an ellipticity for both outer halo profiles of $e = 0.23$, while the ellipticity in the inner part is fitted from the 2MASS data.  This PA and ellipticity are different from the best-fit values in \citet{Rejkuba2022} of PA $= 55^{\circ}$ and $e = 0.46\pm0.02$ -- we note however that their values are not based on azimuthally complete data, but instead are a best-fit to isolated HST fields that sample only a small portion of the outer halo.  While it is possible that the PA twists at large radii, our CJAM modeling cannot accommodate such a twist.  We therefore use the \citet{Dufour1979} values of PA and $e$ throughout the halo.   

We combine the profiles by creating a single image combining the 2MASS data at small radii ($<70\arcsec$), and the \citet{Dufour1979} (for distances between $70-255\arcsec$) and \citet{Rejkuba2022} profiles at large radii (beyond $255\arcsec$).  To do this, the two halo surface density profiles (equation~\ref{eq:dufour_SB} and \ref{eq:Rejkuba_SB}) have to be transformed back into counts using the 2MASS K-band zeropoint ($ZP_{Ks} = 19.915$ mag) to be combined with the 2MASS image. 

\begin{figure}[h!]
    \centering
	\plotone{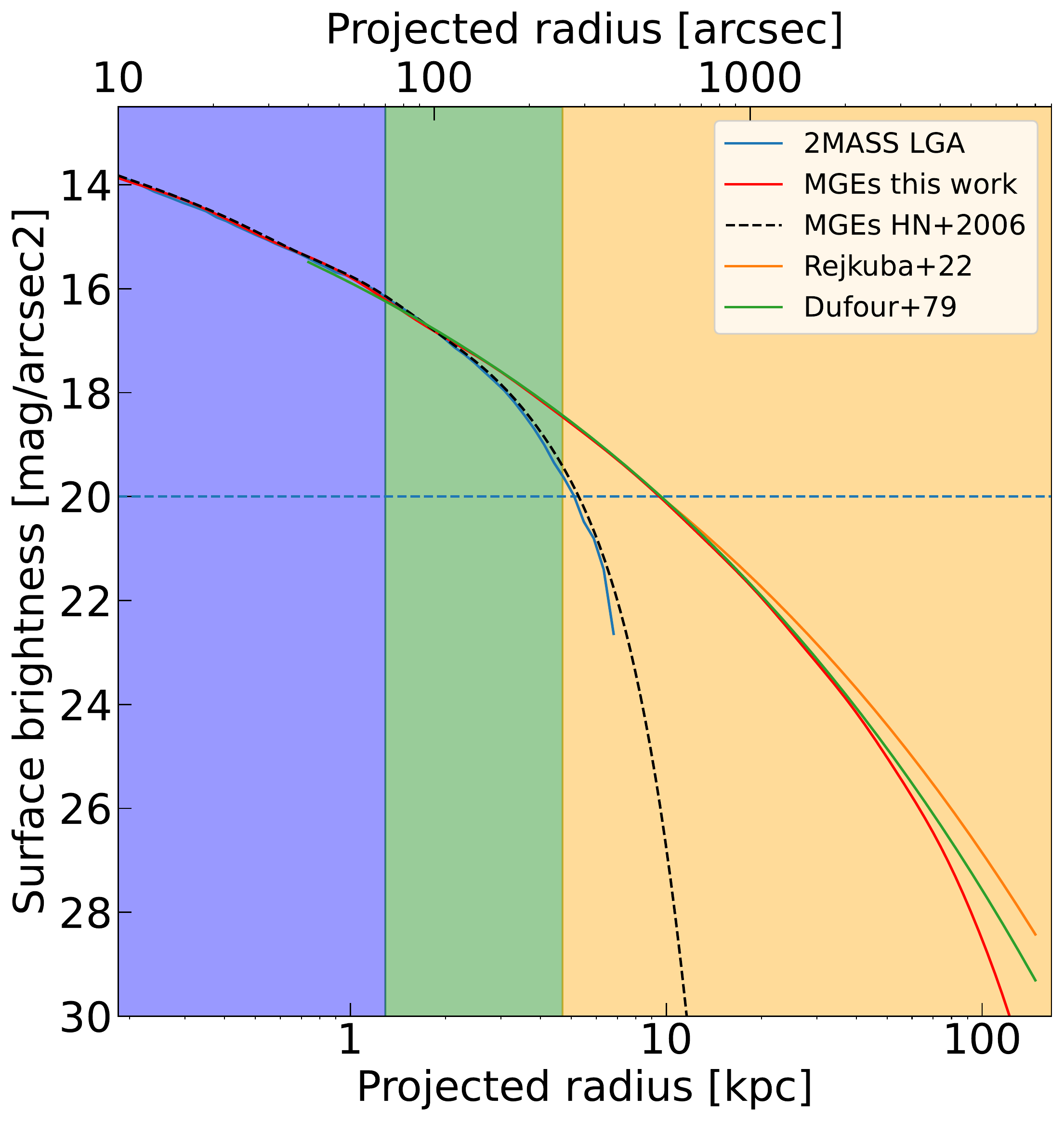}
	\caption{Stellar surface brightness profile of NGC~5128. This figure shows a comparison between the best-fit MGE model for the surface brightness (red) with the 2MASS Large Galaxy Atlas surface brightness profile (blue) \citet{Jarrett2003}, literature MGE model of NGC~5128 (dashed black line) by \citet{Neumayer2006}, and the outer halo profiles in the V-band by \citet{Dufour1979} and \citet{Rejkuba2022} (green \& yellow respectively), assuming a $V-Ks = 3$ mag. The blue horizontal dashed line represents the 2MASS surface brightness $1\sigma$ noise level used to derived the most inner part of the MGE model. The vertical bands show the radius at which we use the 2MASS K-band image, the  de Vaucouleurs-S\'ersic profile by \citet{Dufour1979} and \citet{Rejkuba2022} respectively to derive the MGE model shown in red.}
	\label{fig:surf_profile}
\end{figure}

The resulting combined image is fit using an MGE model, assuming a point-spread function (PSF) of $1.3\arcsec$ determined by measuring the full width at half maximum (FWHM) of stars in the 2MASS image.
%, and a supermassive black hole of $5.5\times10^{7}$ M$_{\odot}$ \citep{Cappellari2009}.  
The resulting final major axis surface brightness profile and the sum of the fitted MGEs are shown in Figure~\ref{fig:surf_profile}, and the MGE parameters are listed in Table~\ref{tab:MGEs}. We note the axis ratios for the MGE components are left as free parameters; since each component has contributions from a range of radii, they do not exactly equal the value assumed for the outer halo in constructing the combined surface brightness profile. We calculate the total K-band luminosity of NGC~5128 by integrating the three-dimensional Gaussian with amplitude, sigma and flattening based on the Luminous MGEs listed in Table~\ref{tab:MGEs}. We find a total K-band luminosity of $1.544\times10^{11}$ L$_{\odot}$ which is in excellent agreement with the results found by \citet{Karachentsev2002}. 

Using this MGE, CJAM can deproject the two-dimensional MGE model into a three-dimensional profile. Multiplied by a mass-to-light ratio, we get a luminous stellar mass density.  

Perhaps surprisingly, there has been no published analysis of the star formation history of the inner few kpc of NGC~5128 based on integral-field spectroscopy or similar methods that would allow us fine-grained constraints on the mass-to-light ratio as a function of radius. However, {Hubble Space Telescope}/WFC3 images for NGC~5128 clearly show the presence of young stars and ongoing star formation inside the dust ring of NGC~5128, which extends out to $\sim 180\arcsec$.  Beyond this radius the light appears to be dominated by an older smoother component extending outwards through the halo.  Because our tracer population is primarily located at large radii, and our interest is in constraining the outer mass profile of the galaxy, we fix the inner (all MGE components with spatial $\sigma \leq 180\arcsec$) to $\Upsilon_{K}= 0.65$ based on results from \citet{Cappellari2009}. They used VLT/{SINFONI} stellar integrated field spectroscopic data in the nuclear regions, and found good agreement with spectroscopy extending out to $\sim$40\arcsec. We then leave the $\Upsilon_{K}$ for the outer halo (i.e., all MGE components with $\sigma$ $>$ $180\arcsec$) as a free parameter in our models. 

\subsubsection{Dark Matter Gravitational Potential}
\label{subsec:dm_potential}
For the dark matter contribution to the gravitational potential we assume a generalized version of the Navarro-Frenk-White (NFW) density profile \citep{Navarro1996, Zhao1996}:

\begin{equation}
    \label{eq:NFW_denstiy}
    \rho_{r} = \frac{\rho_{s}}{(r/r_{s})^{\gamma} (1 + (r/r_{s})^{\eta})^{(3-\gamma)/\eta}}
\end{equation}
with $r_{s}$ being the scale radius in pc, $\rho_{s}$ the scale density in M$_{\odot}$/pc$^{3}$, $\gamma$ and $\eta$ are the inner and outer density slope.  For most models presented here, we assume $\gamma = \eta = 1$, the standard NFW profile from \citet{Navarro1996}.  We also assume a spherical halo, an assumption shared by previous measurements of the dark matter halo of NGC~5128.  While the true shape of the halo is likely somewhat triaxial, the main stellar body of NGC~5128 has a relatively modest ellipticity, and
numerical simulations show that dark matter halos are typically rounder than luminous matter in giant elliptical galaxies \citep[e.g][]{Wu2014}. 

We approximate the NFW profile with an MGE model (extending up to $3\times$ further out than our furthest GC); because the NFW profile is a 3-D density, we obtain the height of the best-fit 1-D Gaussians and multiply these by $\sqrt{2 \pi} \sigma$ to convert these to the equivalent 2-D Gaussians expected by CJAM.

\subsection{Tracer density profiles}
\label{subsec:tracers_density}
Each kinematic tracer has its own projected density profile, and CJAM uses this profile to translate the projected position into a distribution of 3D positions that is needed to estimate their velocity moments.

\subsubsection{Globular Cluster Density Profiles}

Metal-rich (red) and metal-poor (blue) clusters have been typically found to have different spatial and kinematic distributions in both the Milky Way and many other galaxies, with the red GCs being more centrally concentrated and the blue GCs more extended \citep[e.g.][]{Brodie2006}.  For the spatial distributions of these sub-populations, we use the power-law fits to the confirmed blue and red GCs in NGC~5128 from \citet{Hughes2022} fit over a range of 7 to 30\arcmin\ ($\sim 7.7-33$ kpc). They find that the red GC population has a steeper profile with a best-fit power-law slope of $-3.05\pm0.28$, while the blue GC population has a slope of $-2.69\pm0.27$.  

For each GC sub-population we create a radial surface density profile (in units of number/arcmin$^2$)  up to a radius of $5.5^\circ$ ($\sim$ 370 kpc), well beyond the boundary of our observed GC population.  We fit this profile with an MGE model assuming spherical symmetry. The MGE surface density model is then deprojected into a three-dimensional profile (with units of volume density). %and saved as a table. 

\subsubsection{Planetary Nebulae Density Profile}
\label{subsubsec:PNE_profile}
Since PNe are just dead low-mass stars, we  assume they follow the distribution of the main stellar population of NGC~5128. Thus, for their tracer density profile, we just use the MGE surface brightness profile derived in subsection~\ref{subsec:lum_grav_pot}. 

\subsubsection{Satellite Galaxy Density Profile}
The number density profile for dwarfs can be obtained using the whole sample of 42 confirmed dwarf satellite galaxies combining Table 6 from \citet{Crnojevic2019} with Table A.1 from \citet{Muller2021b}.
To avoid binning the data to fit a power-law in such a small sample, we fit instead a power-law to the cumulative probability distribution function of the projected radii of the satellite population \citep[e.g.][]{Carlsten2020}. We are only modeling the kinematics of satellite galaxies at $\leq 250$ kpc and therefore fit the 29 satellite galaxies within this radius (this is higher than the 16 used in our dynamical modeling, since only a subset of these 29 have measured radial velocities). The best-fit cumulative power-law slope is $1.47\pm0.23$, with the uncertainty derived from bootstrapping. A cumulative distribution power-law index of $1.47\pm0.23$ implies a two-dimensional radial density profile (equivalent to the probability distribution function) slope of $-0.53\pm0.23$. This measurement is consistent with the flat surface profiles for dwarf satellite galaxies observed in other nearby galaxies, such as M31 (e.g., \citealt{Doliva-Dolinsky2023}). We use this power-law slope to create a radial surface density profile and fit it with a MGE model that we subsequently deproject into a three-dimensional profile. 

\subsection{Tracer Log-Likelihoods}
\label{subsec:LogLikelihood}
Our \textsc{CJAM} models use the number density MGEs and the gravitational potential MGE to predict the first and second-velocity moments of the kinematic tracers. Using a maximum-likelihood analysis we can compare the CJAM LOS velocity predictions directly with our data. Assuming a Gaussian velocity distribution for the GCs, PNe and dwarf satellite galaxies, a point \textit{i} with coordinates ($x'$,$y'$,$z'$) has a dynamical probability of:

\begin{equation}
    P_{dyn,i} = \frac{1}{\sqrt{\sigma_{i}^{2} + \delta v_{z',i}^{2}}} exp[-\frac{1}{2}\frac{(\delta v_{z',i} - \mu_{i} )^{2}}{\sigma_{i}^{2} + \delta v_{z',i}^{2} }]
\end{equation}

\noindent Where $\mu_{i}$ and $\sigma_{i}$ are the LOS mean velocity ($vc$ from \textsc{CJAM}) and velocity dispersion ($\sqrt{|v2zz|}$ from \textsc{CJAM}) predicted by our \textsc{CJAM} dynamical models.

Since the GCs, PNe and dwarf satellite galaxies are independent tracers (but modeled with the same gravitational potential), each population has separate probabilities, giving a total log-likelihood:

\begin{equation}
\begin{split}
\noindent
    \ln \mathcal{L} = & \Sigma_{i=1}^{N_{GC,red}}\ln P_{dyn,i}^{GC,red} + \Sigma_{i=1}^{N_{GC,blue}}\ln P_{dyn,i}^{GC,blue} + \\ 
    & \Sigma_{i=1}^{N_{PNe}}\ln P_{dyn,i}^{PNe}  + \Sigma_{i=1}^{N_{Dwarfs}}\ln P_{dyn,i}^{Dwarfs}
    %= \mathcal{L}_{GC} + \mathcal{L}_{PNe} + \mathcal{L}_{Dwarfs}
\end{split}
\end{equation}

\noindent where $GC$, $PNe$, and $Dwarfs$ are used for our globular cluster, planetary nebulae and dwarf satellite galaxy tracers.

\subsection{Markov Chain Monte Carlo Optimization}
\label{subsec:MCMC_optimization}
To effectively explore the parameter space of our discrete dynamical models we use a MCMC implementation. We use the \textsc{Python} software \textsc{emcee} \citep{EMCEE} for an affine-invariant MCMC ensemble sampler.

For the proposed distribution of the walkers we used a combination of two moves, ``DESnookerMove'' with an $80\%$ weight and ``DEMove'' with a $20\%$ weight, effectively meaning that the walkers will choose $80\%$ of the times the ``DESnookerMove'' proposal and $20\%$ of the times the ``DEMove'' proposal. This combination of proposals for the walkers is suggested in the documentation of \textsc{emcee} for ``lightly'' multimodal model parameters (more details about the ``DESnookerMove'' and ``DEMove'' can be found in the references within the \textsc{emcee} documentation\footnote{\tt https://emcee.readthedocs.io/en/stable/}). We use this mixture of walker proposals as it has already been tested for multimodal parameters and we found it improves the convergence of our models relative to e.g.~the default ``stretch move". 

For each model, we use the double amount of walkers as the number of free parameters in the model, and run \textsc{emcee} for 1000 steps. Based on analysis of the auto-correlation functions, we discard the first 200 steps as burn in and use the remaining 800 steps to characterize the post-burn distribution of the free parameters. 

\subsection{Fiducial Dark Matter Halo Model}
\label{subsec:Fiducial_model}

For the Fiducial model (capitalized for clarity throughout), we use the full sample of tracers as discussed in Section~\ref{sec:Kinematic_Tracers}.

The model assumes each tracer has its own anisotropic orbital distribution.  For the dark matter halo, we assume an NFW profile for the dark matter component.  In total, there are seven free parameters:

\begin{enumerate}
    \item $K$-band stellar mass-to-light ratio ($\Upsilon_{K}$) outside a radius of 180\arcsec\ (see additional discussion below).
    \item Logarithm of the dark matter scale density, $\log(\rho_{s})$.
    \item $\log(Ds) \equiv \log(\rho_{s} r_{s}^{3})$. This is a proxy for the dark matter scale radius $r_s$ that is the parameter of interest, but fitting for $\log(Ds)$ rather than r$_{s}$ reduces the degeneracy between $\rho_{s}$ and $r_{s}$.
    \item Four $\beta$ velocity anisotropy parameters, one for each different tracer: red GCs ($\beta_{red}$), blue GCs ($\beta_{blue}$), PNe ($\beta_{PNe}$), and dwarf satellite galaxies ($\beta_{dwarf}$). 
\end{enumerate}
As mentioned in Section~\ref{subsec:lum_grav_pot}, we fix $\Upsilon_{K}=0.65$ for the inner $180\arcsec$, based on results of \citet{Cappellari2009}. However, we leave $\Upsilon_{K}$ as a free parameter for radius beyond $180\arcsec$ with a Gaussian prior of 1.08$\pm$0.30 based on the distribution of dynamical stellar $\Upsilon_{K}$ of early-type galaxies 
from the ATLAS$^{3D}$ project \citep{Atlas3Dcappellari}. 

For the free parameters in the NFW profile, we use a Gaussian prior for their corresponding virial concentration $c_{vir} = r_{vir}/r_{s}$, where $r_{vir}$ is the virial radius (see Section~\ref{subsec:Fiducial_results}). We set this Gaussian prior using the relation between concentration and the virial mass $c_{vir} = 15 - 3.3\log({\rm M}_{vir}/10^{12}h^{-1}  {\rm M}_{\odot})$ from the simulations of \citet{Bullock2001}. With an assumed M$_{vir} = 1.3\times 10^{12} {\rm M}_{\odot}$ for NGC~5128 from \citet{Woodley2007}, we get a mean concentration $c_{vir}=14.0$ with a standard deviation of $5.6$ calculated from the $40\%$ scatter in the \citet{Bullock2001} relation. Anticipating our final results, we find a higher virial mass (and hence lower expected $c_{vir}$) for NGC~5128, but the broad nature of this prior allows the likelihood to dominate in our modeling.

We use a uniform prior from $-$0.4 to 0.4 for the anistropy parameters at all radii for $\beta_{red}$, $\beta_{blue}$, $\beta_{dwarf}$.  We use this same prior for $\beta_{PNe}$, but this parameter only applies outside of 180\arcsec.  Inside  $180\arcsec$,
we set $\beta_{PNe}$ to the published radial velocity anistropy profile from the stellar kinematic fits of \citet{Cappellari2009} ($\beta_{PNe}=-0.35$ for $R\leq 0.5\arcsec$, $\beta_{PNe}=0.2$ for $0.5\arcsec<R\leq10\arcsec$, and $\beta_{PNe}=-0.35$ for $10\arcsec<R\leq180\arcsec$). We verify that this anisotropy profile provides a good fit to the central stellar kinematics described in Section~\ref{sec:Kinematic_Tracers}.

\section{Dynamical Modeling Results}
\label{sec:Model_results}
In this section, we fit our Fiducial dynamical model to the data and summarize the results (Section \ref{subsec:Fiducial_results}). We then 
explore the sensitivity of these results by varying the data fit (Section \ref{subsec:examining_uncertainties_data}) and different model assumptions (Section \ref{subsec:examining_uncertainties}).

\subsection{Fiducial Model Results}
\label{subsec:Fiducial_results}
The MCMC implementation of the Fiducial model consists of 14 walkers with 1000 steps. The MCMC converges after 200 steps, and the post-burn probability distribution functions (PDFs) for each parameter are shown in the left side of Figure~\ref{fig:Fiducia_corner_plot} and Table~\ref{tab:Best_fit_values}.

Most of the parameters are well-constrained by the data. We observe medium to strong correlations between the stellar mass-to-light ratio $\Upsilon_{K}$ (which determines the mass in stars) and the NFW profile parameters. Such correlations are expected: our dynamical tracers constrain the total mass, and thus a more massive luminous component will require a less massive dark matter halo component (and vice versa).

While the use of $\log(Ds) = \log(\rho_{s} r_{s}^{3})$ as a proxy for the NFW scale radius $r_{s}$ partially helps break the degeneracy between $\rho_{s}$ and $r_{s}$ in the fits, we still observe a negative correlation between $\log(Ds)$ and $\log(\rho_{s})$. This is also expected given that both parameters contain $\rho_{s}$. In any case, both NFW halo parameters are well-constrained by the data.   All parameter estimates quoted below are calculated as the median of the posterior PDF, with the listed uncertainties corresponding to $1\sigma$ equivalent (68\%) of the PDF around the median.

\subsubsection{Dark Matter Halo Parameters}

\begin{figure*}[h!]
    \centering %\plottwo{Figures/Corner_plot_Fiducial.pdf}{Figures/Corner_plot_physical.pdf}
    \plottwo{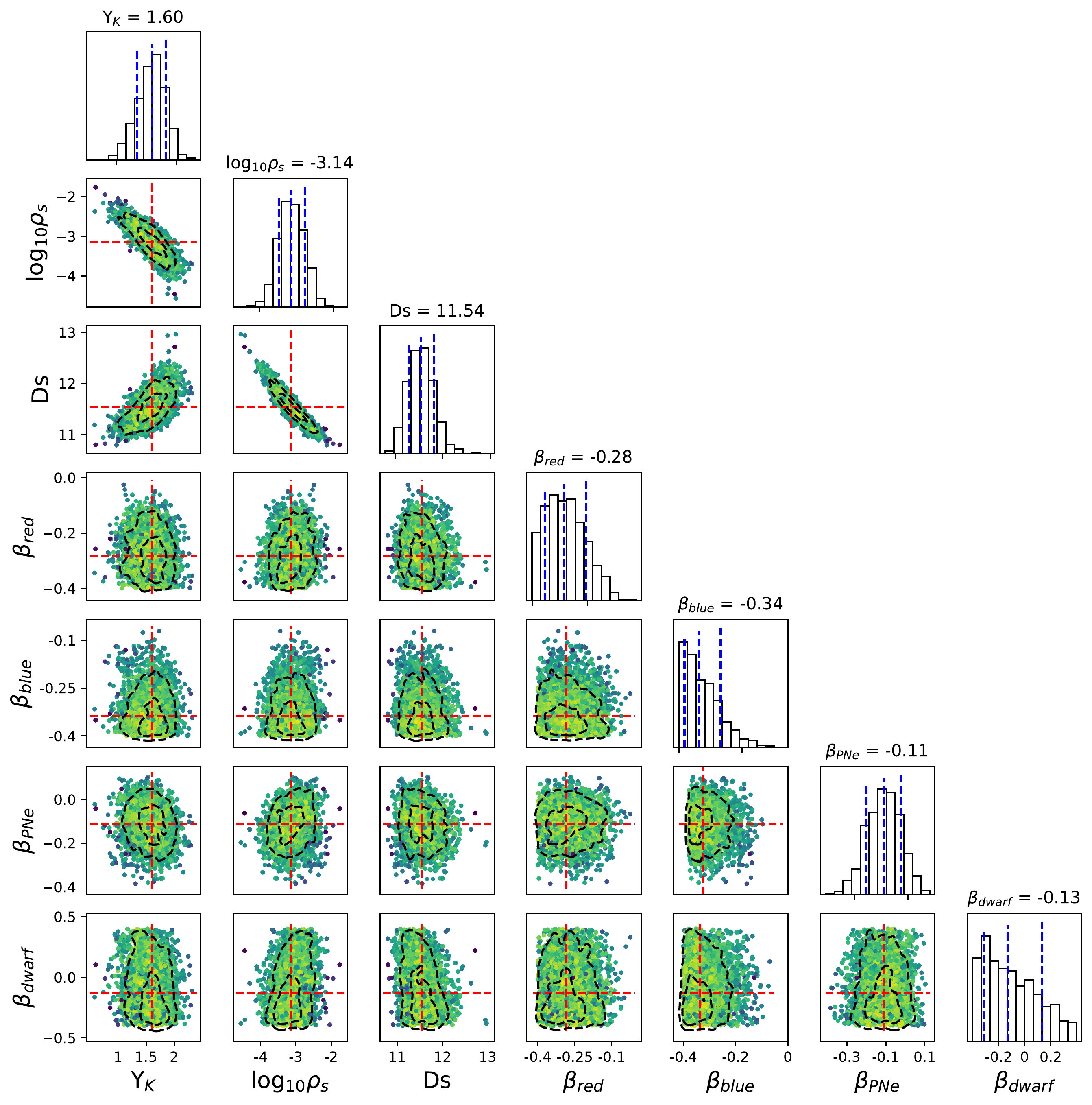}{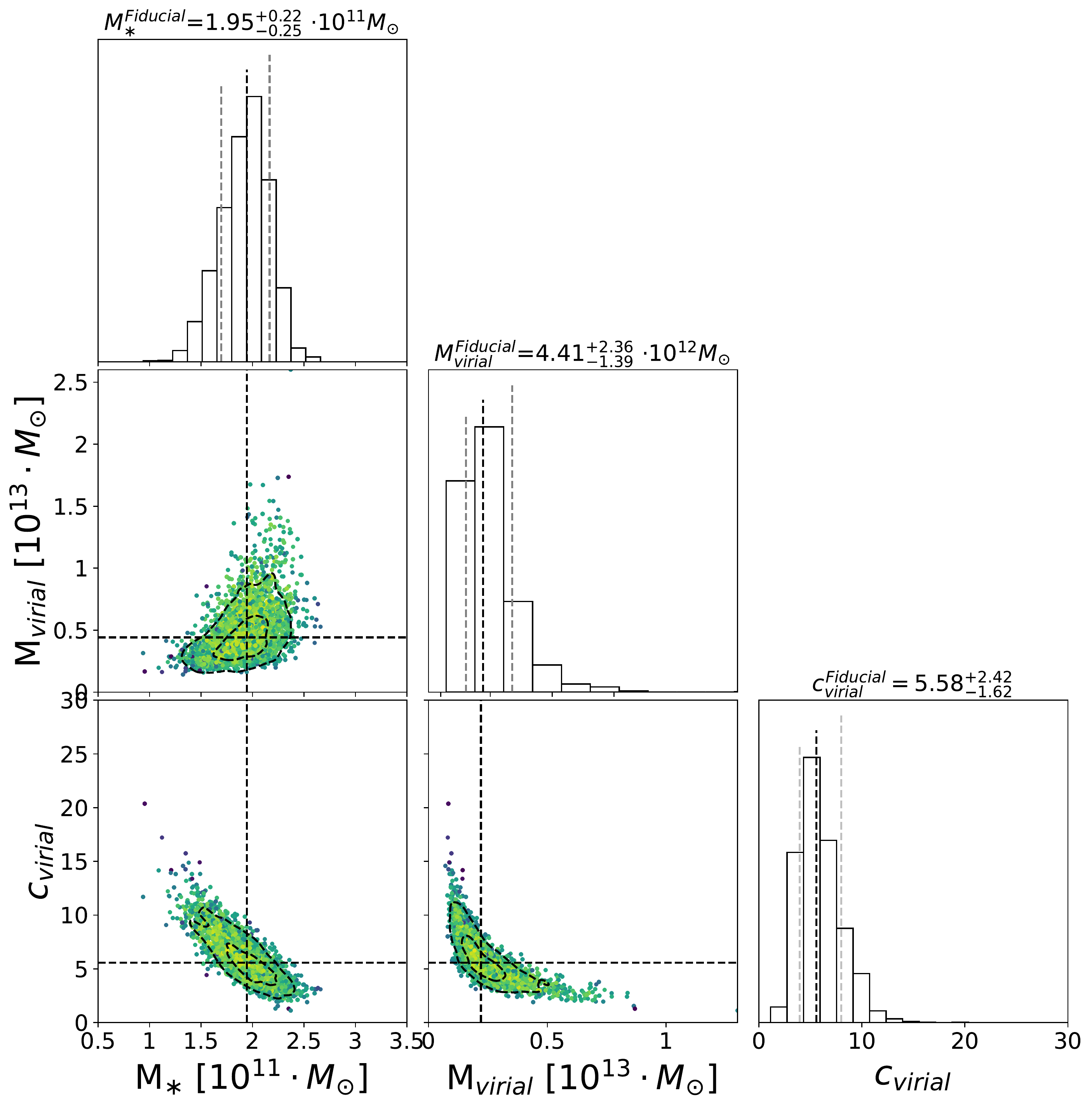}
    \caption{\textit{Left:} MCMC post-burn distribution for the seven free parameter of the Fiducial model. The histograms show their marginalized PDF, and the vertical blue lines show their best-fit value and their 1-$\sigma$ error. The scatter plots show the two-dimensional PDF for the parameters colored by log-likelihood, with yellow colors indicating higher values. The red-dashed lines indicate each parameter's best-fit value (median) location, and the black-dashed contours show their 1-$\sigma$ and 2-$\sigma$ error. \textit{Right:} Plots show the one and two-dimensional PDF distribution for ${\rm M}_{vir}$, ${\rm M}_{\ast}$ and $c_{vir}$. Lines and contours in this plot are the same as in the left panel, but all are shown here in black.}\label{fig:Fiducia_corner_plot}
\end{figure*}

For the NFW scale parameters, we find a $\log(\rho_{s}) = -3.14^{+0.37}_{-0.33}$ and $r_{s} = 78^{+45}_{-29}$ kpc, with the latter derived from $\log(Ds)$. To more directly compare our results with published values and with simulations, we calculated the corresponding virial mass (${\rm M}_{vir}$), virial concentration ($c_{vir}$), and luminous mass (${\rm M}_{\ast}$) based on our best-fit parameters shown in Figure~\ref{fig:Fiducia_corner_plot} (left). We show these translations for each of our MCMC walkers in Figure~\ref{fig:Fiducia_corner_plot} (right) and Table~\ref{tab:derived_quantities}. For further information on how we translated our fitted parameters to the inferred dark matter halo model parameters, please refer to Appendix~\ref{appendix}. 

Our Fiducial model fit has a dark matter halo virial mass of ${\rm M}_{vir}=4.4^{+2.4}_{-1.4}\times10^{12} {\rm M}_{\odot}$, a concentration of $c_{vir}=5.6^{+2.4}_{-1.6}$, and a corresponding virial radius of $r_{vir}=432^{+66}_{-51}$ kpc. Using the best-fit $\Upsilon_K$ (see Section~\ref{sec:MLfit}), the total baryonic mass of the Fiducial model is $1.9^{+0.2}_{-0.3}\times10^{11} {\rm M}_{\odot}$. This gives us a total mass (${\rm M}_{vir,t} = {\rm M}_{vir} + {\rm M}_{bary}$) of ${\rm M}_{vir,t} = 4.7^{+1.4}_{-1.4}\times10^{12} {\rm M}_{\odot}$. To facilitate comparison with other studies, we also calculated other commonly used quantities referenced to a cosmological overdensity of $\Delta = 200$. Our Fiducial model gives ${\rm M}_{200} = 3.6^{+1.7}_{-1.0} \times10^{12} {\rm M}_{\odot}$ and $c_{200} = 4.2^{+1.9}_{-1.3}$, corresponding to a total mass of ${\rm M}_{200,t} = 3.8^{+1.7}_{-1.0}\times10^{12} {\rm M}_{\odot}$. These quantities are smaller than the virial ones; at $z=0$, ${\rm M}_{200}\sim0.8 {\rm M}_{vir}$.  
The Fiducial model has a dark matter fraction $f_{DM}$ of $0.11\pm0.04$ 
within $1r_{eff}$ and $0.52\pm0.08$ within $5r_{eff}$.

\subsubsection{Mass-to-Light Ratio}
\label{sec:MLfit}

We determined a best-fit value of $\Upsilon_{K} = 1.60^{+0.22}_{-0.25}$  for the mass-to-light ratio in the outer region of NGC~5128 ($>180\arcsec \:; \gtrsim 3.3$ kpc). This value is about 2.5 times larger than the $\Upsilon_{K}=0.65$ found by \citet{Cappellari2009} for the inner regions of the galaxy ($\lesssim 0.7$~kpc), and more than 1$\sigma$ larger than the prior value of $\Upsilon_{K} = 1.08\pm0.3$ assumed in our dynamical models (see Subsection~\ref{subsec:Fiducial_model}).

Given that we expect the outer stellar regions of NGC~5128 to be older than the inner regions, it is not surprising that the $\Upsilon_{K}$ value fit in the outer regions is larger. However, it also lies at the upper edge of the values favored by our prior, perhaps suggesting a tension between the data and the prior.
There are several potential solutions to this discrepancy. First, simulations conducted by \citet{Li2016} suggest that JAM dynamical models tend to overestimate the baryonic mass and underestimate the inner dark matter content in triaxial galaxies. Since some previous studies found evidence that NGC~5128 is triaxial \citep{Hui1995,Peng2004}, this could in principle explain the overestimation of the baryonic mass and hence a too-high $\Upsilon_{K}$. Another, perhaps more likely possibility is that the tracers in the inner, more baryon-dominated regions are not in virial equilibrium due to the residual effects of a major merger (\citealt{Wang2020}; see additional discussion in Section \ref{subsec:examining_uncertainties_data}).

\subsubsection{Tracer Anisotropy}
\label{subsec:trac}

In our model fits, all tracers nominally exhibit tangential velocity anisotropy, but this is statistically significant only for the red and blue GCs, both of which have $\beta \sim -0.3$. The PNe are very mildly anisotropic ($\beta \sim -0.11\pm0.08$) and the dwarf satellite galaxies are close to isotropic, though the anisotropy is poorly constrained for the latter due to small numbers. We provide more detailed analysis of the velocity anisotropy profiles for each tracer in Section~\ref{radvar}.
 
We note that the observed appearance of velocity anisotropy depends on inclination. As in previous dynamical modeling of both NGC~5128 \citep[e.g][]{Hui1995,Peng2004} and other galaxies \citep[e.g][]{Zhu2016}, we assume an edge-on inclination in our modeling. We investigate this assumption in more detail in Section~\ref{subsec:inclination}.

A caveat on the precise values of anisotropy inferred is that  previous studies have shown significant rotation for both the PNe \citep{Peng2004,Woodley2007,Coccato2013} and the metal-poor and metal-rich GCs \citep{Woodley2007,Woodley2010,Coccato2013,Hughes2022}. We experimented with the inclusion of rotation in our models, but in the end chose not to simultaneously model both anisotropy and rotation due to the computational expense. While this is not expected to affect the mass estimates, the inclusion of rotation in the models could change the inferred anisotropy for the tracers. This is an interesting potential degeneracy to investigate in future work.

\subsubsection{Velocity Dispersion Profiles: Comparing Model and Data}

\begin{figure}
    \centering
    \plotone{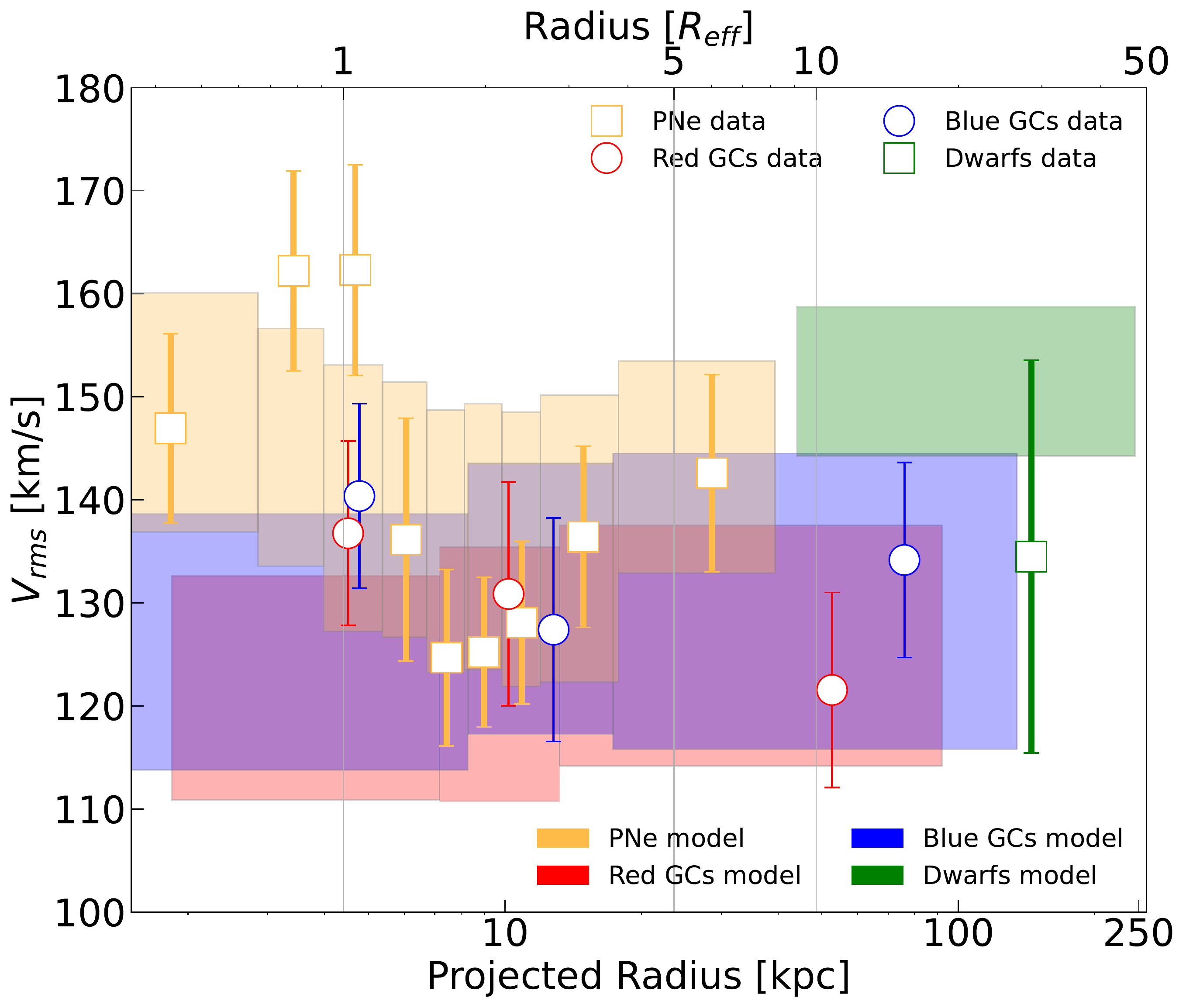}
    \caption{Comparison of observed $V_{rms}$ radial profiles for the different kinematic tracers (outlined points) to predicted profiles from our best-fit Fiducial dynamical model (colored rectangles). The colors represent the same tracers as in Figure~\ref{fig:Tumpet plot}. This figure shows the Fiducial model does an excellent job of explaining the overall velocity distribution of our tracers.}
    %The squares represent the binned derived $V_{rms}$ for each tracer following the iterative method from \citet{Walker2009}. 
    \label{fig:rotation_model}
\end{figure}

In Figure~\ref{fig:rotation_model} we present a visual comparison of our Fiducial model to the tracer observations. Focusing first on the data, the outlined white points represent the observed tracer velocity dispersion measurements binned using a maximum likelihood estimator from \citet{Walker2009} with errors calculated using bootstrapping.  This velocity dispersion is equivalent to the observed $V_{rms}$, since it includes no rotation component and is azimuthally averaged.
The $V_{rms}$ for the red and blue GCs remains mostly constant with radius with values around $V_{rms} \approx 120$--140 km s$^{-1}$. For the PNe, $V_{rms}$ is highest in the center at 
$\approx 160$ km s$^{-1}$, but shows an abrupt drop at around $1 r_{eff}$ and non-monotonic variations at larger radii. These features may be related to the possibilities of non-virial motions or triaxiality discussed above. 

To compare these measurements to the model, we derive predicted $V_{rms}$ profiles for our Fiducial model for each of the individual tracers. The model values were evaluated at the same points as the data in each radial bin.  The models are shown as colored rectangles, with the height on these rectangles indicating the standard deviation in the modeled $V_{rms}$ values due to the radial and azimuthal variation of data points within each bin.   We note that the binning in Figure~\ref{fig:rotation_model} is solely for the purposes of this comparison---all dynamical modeling fits were made to unbinned data.

We generally find good agreement between the data and model predictions in Figure~\ref{fig:rotation_model}, with the shaded data regions for each tracer consistent with model predictions.   Hence our NGC~5128 mass model appears to be a reasonable fit to the observed tracer populations.

In the next two subsections we explore the robustness of our model fits by exploring their dependence on (i) the details of the datasets used and (ii) the model assumptions.

\subsection{Checking Robustness: Data Variations}
\label{subsec:examining_uncertainties_data}

\begin{figure}[h!]
    \centering
    \plotone{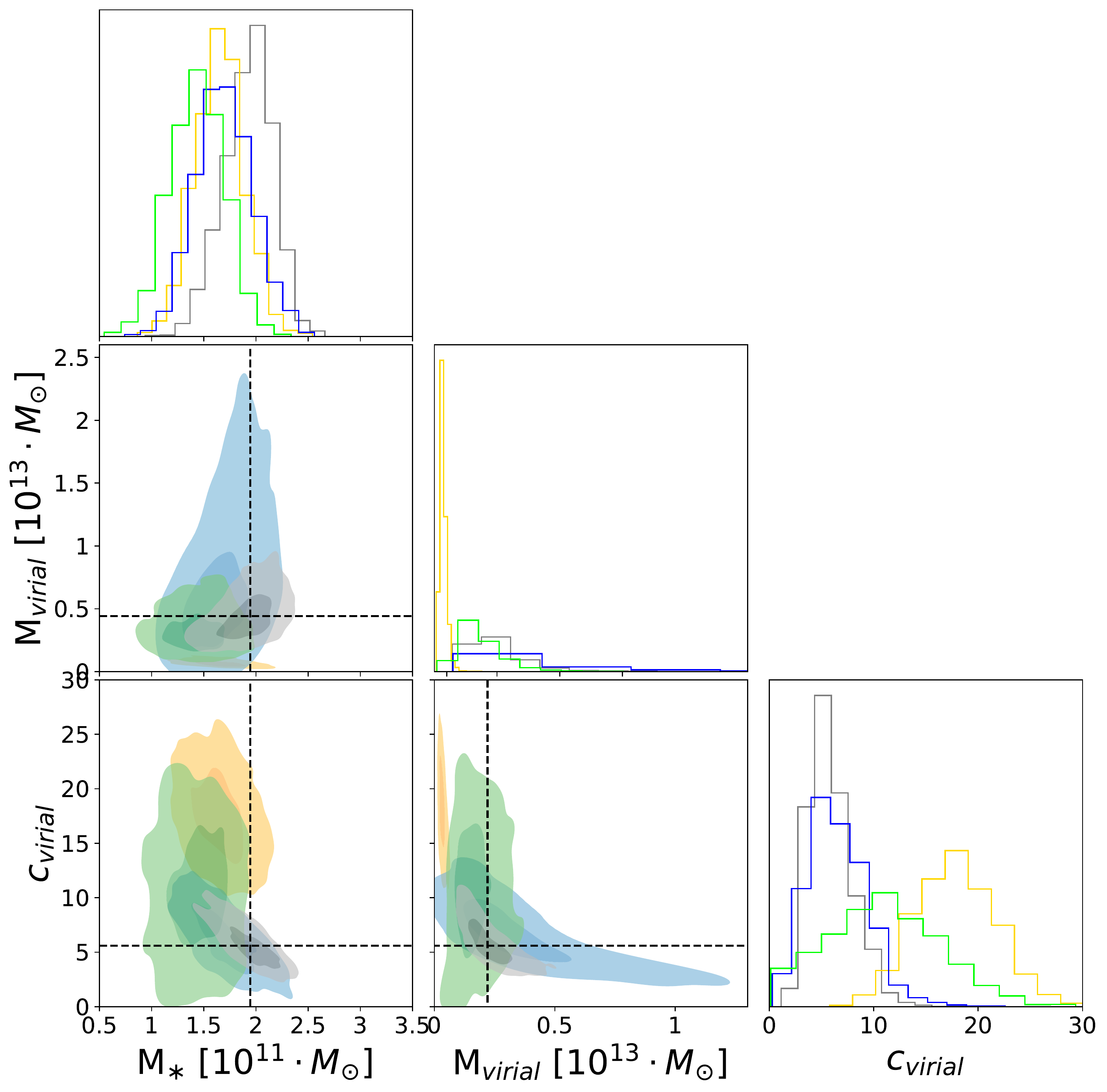}
    \caption{ One and two-dimensional PDF of ${\rm M}_{vir}$, $c_{vir}$, and ${\rm M}_{\ast}$ of the Fiducial model compared to the PDF inferred from single-tracer anisotropic models. The blue PDF represents the GC-only model, the yellow PDF the PNe-only model, and the green PDF the dwarfs-only model. We represent  1-$\sigma$ and 2-$\sigma$ confidence intervals for each model with darker and lighter colors, respectively. The dashed lines show the median value for each quantity for the Fiducial model, and are the same as in Figure~\ref{fig:Fiducia_corner_plot} right. }
    \label{fig:Fiducial_vs_tracers}
\end{figure}

We begin by separating the classes of tracers (GCs, PNe, and dwarfs) and fitting the same class of model as for our Fiducial model: a stellar + anisotropic NFW halo model, fit to each class of tracer individually. The results of these fits are listed in  Table~\ref{tab:Best_fit_values} and plotted in Figure~\ref{fig:Fiducial_vs_tracers}.

Focusing first on the inferred halo parameters, the concentration $c_{vir}$ and virial mass ${\rm M}_{vir}$ measured from the GCs are consistent with the Fiducial model within the uncertainties. The same is true for the dwarfs, though the dwarfs $c_{vir}$ is poorly constrained owing to small numbers and their restricted radial range. However, the derived halo parameters from the PNe are drastically different than the Fiducial model, with a much more concentrated and less massive halo ($c_{vir} = 18\pm4$; ${\rm M}_{vir} = 0.7^{+0.3}_{-0.2} \times 10^{12} {\rm M}_{\odot}$).

This tension does not appear to be driven by differences in the inferred anisotropy for the tracers. The anisotropy of the red and blue GCs were essentially identical to those in the Fiducial model, while that of the PNe is marginally more negative, but not to a sufficient degree to explain this discrepancy. In Section \ref{radvar} below we explore whether allowing radial variations in the anisotropy could relieve the tension between GCs and PNe. Anticipating these results, we do not find significant observational support for such variations.

The stellar masses (${\rm M}_{\ast}$) inferred from the separate tracer fits all agree within $1\sigma$ with the Fiducial model and with each other. However, the three single-tracer models find a smaller $\Upsilon_{K}$ than the Fiducial model, yielding less mass in stars and more in dark matter at these inner radii. This is the expected degeneracy between the apportionment of mass between stars and dark matter, and the advantage of the Fiducial model is that using a range of tracers with different radial distributions helps to partially break this degeneracy.

As mentioned above, several previous papers have shown that the PNe display complex kinematics, with minor axis rotation at small radii and major axis rotation at larger radii. \citet{Peng2004} suggested this could be due to triaxiality. Axisymmetric JAM modeling of triaxial systems has been studied by \citet{Li2016}; they find that the impact compared to spherical models is only about 20\% in mass, much lower than the discrepancies we see between the PNe and GCs.  A more straightforward interpretation of these results is that at least a subset of the NGC~5128 PNe are not virialized, likely due to the recent merger \citep{Israel1998,Wang2020}, consistent with the findings of \citet{Coccato2013}. 

Overall, the single tracer fits show that the GCs and dwarfs give mutually consistent results that also agree with the Fiducial model, while the PNe strongly differ. As a cross-check of the effect of the PNe on the fits, we also performed a fit that used the GCs, dwarfs, and the inner ($< 1r_{eff}$) PNe, excluding the outer PNe that show unusual rotation around the major axis \citep{Peng2004}. This fit gave halo parameters, the stellar mass-to-light ratio, and tracer anisotropies all consistent with the Fiducial model within the uncertainties, again supporting the robustness of our Fiducial model fitting results even in the presence of potential non-virial motions for a subset of the tracers.

\subsection{Checking Robustness: Model Variations}
%\subsection{Examining Systematics through Model Variation}
\label{subsec:examining_uncertainties}

\subsubsection{Comparison to Isotropic Models}
\label{subsec:Fiducial_vs_Isotropic}

The results of the previous two sections show consistent support for tangential anisotropy in all the tracer populations, with the strongest evidence for the GCs and some evidence for the PNe.
Since there is a well-known degeneracy between mass and anisotropy in dynamical models (e.g., \citealt{Binney1982}), and studies often assume isotropy for simplicity, here we fit isotropic models to allow a comparison to our Fiducial models.

\begin{figure}
    \centering
    \plotone{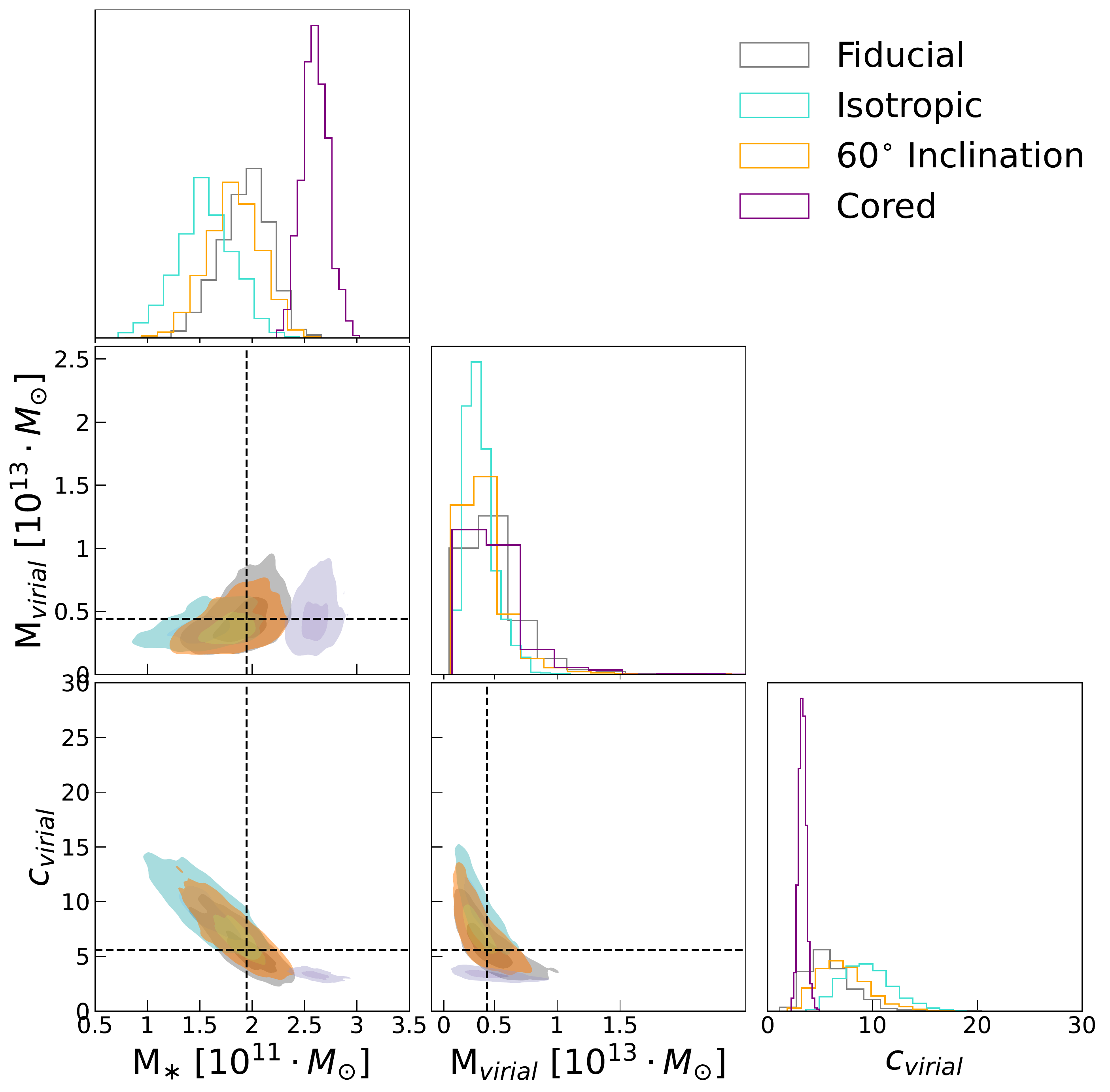}
    \caption{One and two-dimensional PDF distribution of the ${\rm M}_{vir}$, ${\rm M}_{\ast}$, and $c_{vir}$ for the Fiducial model compared to the model variations discussed in Section~\ref{subsec:examining_uncertainties}. The PDF distributions for each model are shown in different colors: black for the Fiducial model, cyan for the "Isotropic model," orange for the "$60^{\circ}$ inclination model", and fuchsia for the "cored" model. The darker and lighter colors represent the 1 and $2-\sigma$ confidence intervals, respectively, for each model. The median value for the Fiducial model are shown in dashed black lines, and listed in Table~\ref{tab:derived_quantities} for the rest of the models.}
    \label{fig:Fiducial_vs_model_variations}
\end{figure}

The results of these fits are shown in Figure~\ref{fig:Fiducial_vs_model_variations} and listed in Table~\ref{tab:Best_fit_values}. ${\rm M}_{vir}$ for the isotropic model is $3.7^{+1.4}_{-0.9} \times 10^{12} {\rm M}_{\odot}$, about 15\% lower than the Fiducial model but consistent within the uncertainties.
The inferred stellar mass is lower by a similar amount, leading to a somewhat higher inferred concentration $c_{vir} = 9.3^{+2.7}_{-2.2}$ that is nevertheless formally consistent with the Fiducial model value of 
$c_{vir}$=5.6$^{+2.4}_{-1.6}$. The isotropic model has a dark matter fraction $f_{DM}$ of $0.19^{+0.07}_{-0.05}$ within $1R_{eff}$ and $0.66^{+0.06}_{-0.08}$ within $5R_{eff}$.

\subsubsection{Radially Varying Anisotropy}
\label{radvar}

Here we briefly explore whether the assumption of radially-constant anisotropy for each tracer is meaningfully affecting our results.
We fix the gravitational potential parameters to the values of the Fiducial model and then run single-tracer fits, but replacing a single value of $\beta$ with four logarithmically spaced radial ranges where it can take on different values in each region.
We find no statistically significant evidence for a radial gradient in $\beta$ for any of the tracer populations, indicating the assumption of a radially-constant velocity anisotropy for each tracer is reasonable, at least for this dataset.

\subsubsection{Varying Inclination}
\label{subsec:inclination}
In this subsection we study the effect of the assumed inclination in our dynamical models. %Precious 
As mentioned in Section \ref{sec:Discrete_model}, previous studies suggest the inclination of NGC~5128 is more likely edge-on, and in particular that face-on orientations are disfavored \citep{Hui1995,Peng2004}.

Here we compare our edge-on ($90^{\circ}$) model to one where the assumed inclination is instead the median value for a random distribution: $60^{\circ}$. The other model assumptions are identical to those of the Fiducial model. The resulting ${\rm M}_{vir}$, $c_{vir}$, and ${\rm M}_{\ast}$ are shown in Figure~\ref{fig:Fiducial_vs_model_variations}, and listed in Table~\ref{tab:Best_fit_values}.

The model with an inclination of $60^{\circ}$ produces results consistent with the Fiducial model to within 1$\sigma$ for all parameters, with a slightly higher concentration and a slightly lower (about 15\% lower) virial mass.

\subsubsection{A Cored Halo}

For our Fiducial model we have assumed a standard cuspy NFW profile, corresponding to $\gamma = \eta =1$ in Equation \ref{eq:NFW_denstiy}. While giant elliptical galaxies tend to have dark matter profiles well-fit by a standard NFW profile \citep[e.g][]{Grillo2012}, here we briefly consider the sensitivity our results to the assumed halo profile.

As a comparison ``cored" model, we assume $\gamma=0$ and $\eta = 2 $ in Equation \ref{eq:NFW_denstiy}, producing a central core rather than a cusp. All other model parameters are the same. 

The differences in the fitted parameters between the cored and Fiducial are generally as one would expect: ${\rm M}_{vir}$ is essentially identical, being set by the data at larger radii, while the concentration is lower ($c_{vir} = 3.3\pm0.4$) due to the lesser central dark matter density. This forces the stellar mass-to-light ratio an implausibly high value $\Upsilon_{K} = 2.3\pm0.1$ to match the inner kinematics, compensating for the lower dark matter density. This is in strong tension with our prior on $\Upsilon_{K}$ ($1.08\pm0.30$), suggesting this cored dark matter model is a poor fit compared to the standard NFW profile.

\subsection{Conclusions}

Our main conclusions from this section are:
\begin{itemize}
\item Our Fiducial NFW model provides a good fit to nearly all the data, potentially with the exception of a subset of the PNe.
\item The results are insensitive to modest variations in the tracer population used.
\item The results are also relatively insensitive to modest variations in model assumptions about anisotropy or inclination, and such variations are not inconsistent with the data. However, a cored dark matter halo is disfavored.
\end{itemize}

Again, we summarize our mass modeling results in Table~\ref{tab:Best_fit_values}.  We next compare our results with previous mass and mass profile measurements of NGC~5128.

\section{Comparisons to Previous NGC~5128 Total Mass and Halo Estimates}
\label{sec:compare}

\subsection{Comparison to Enclosed Mass Estimates}
\label{subsec:comparison_to_mass}

\begin{figure*}
    \centering
    \begin{minipage}[c]{0.4\linewidth}
        \centering
        \includegraphics[width=1.20\linewidth]{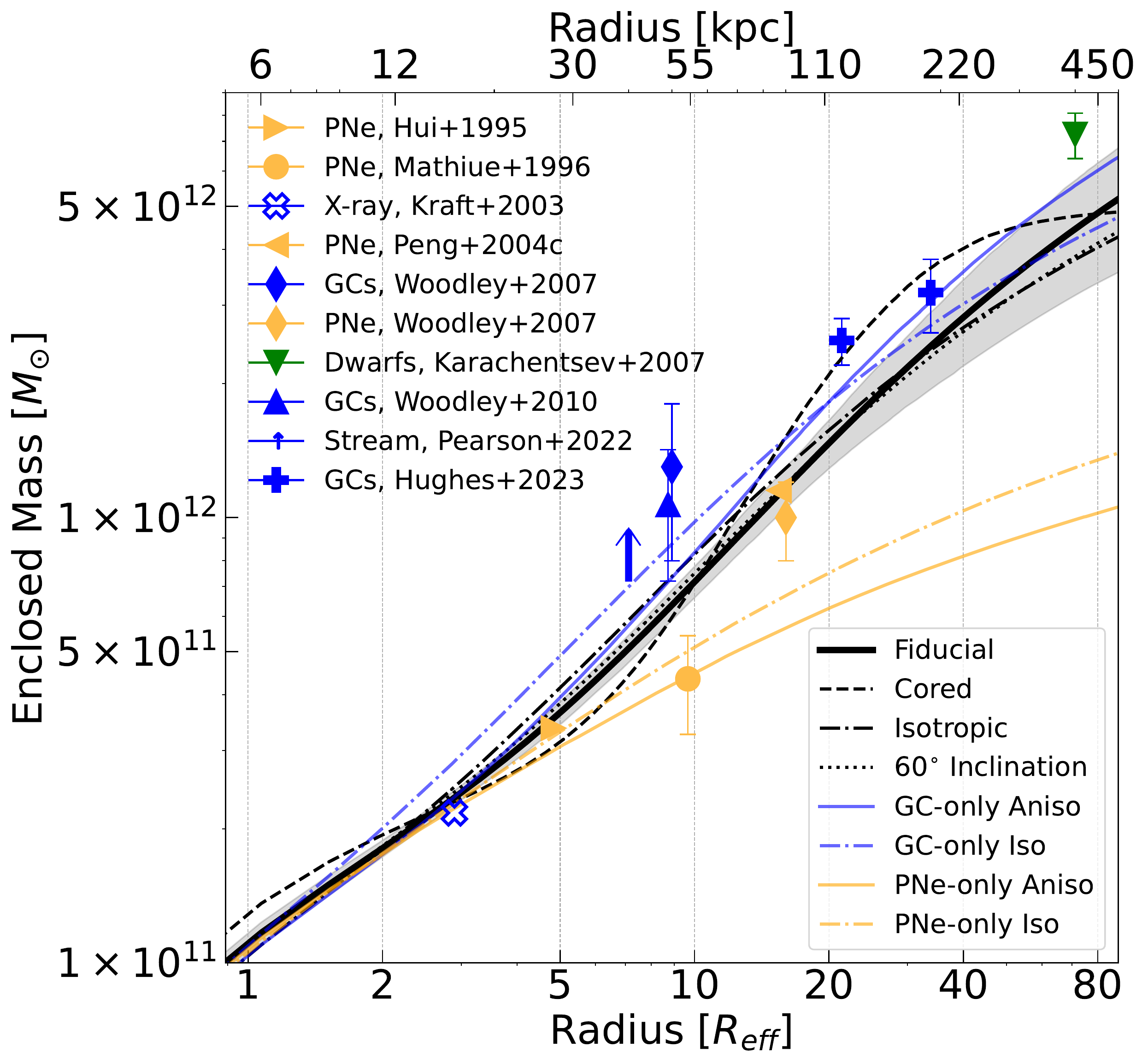}
    \end{minipage}
    \hspace{15mm}
    \begin{minipage}[c]{0.4\linewidth}
        \centering
        \includegraphics[width=1.11\linewidth]{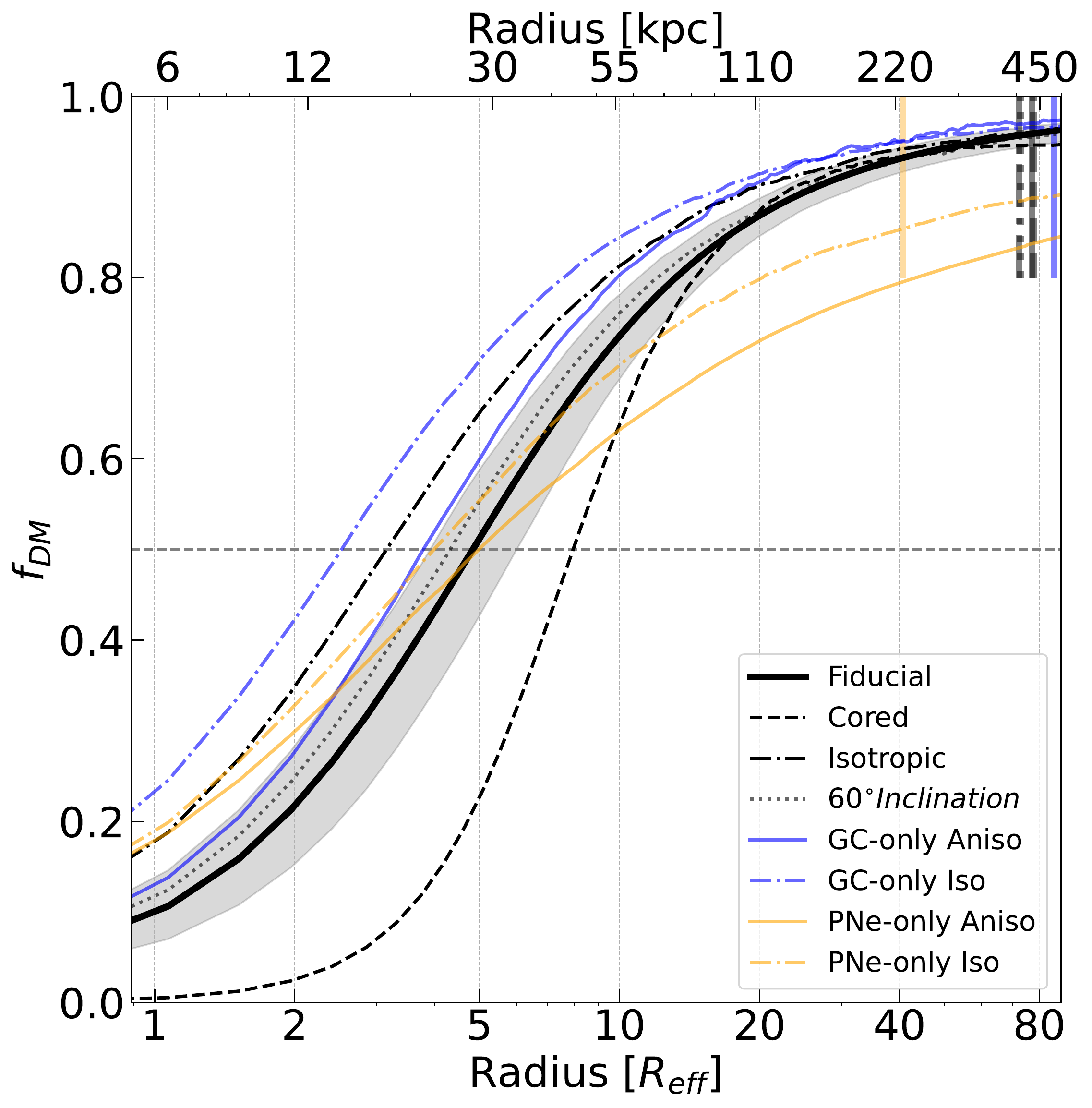}
    \end{minipage}%
    %\plottwo{Figures/Enclosed_mass.pdf}{Figures/DM_fraction.pdf}
    \caption{\textit{Left:} Enclosed mass for the Fiducial model as a function of galactocentric radius compared to literature measurements and different dynamical models from this work. The solid lines represent the median enclosed mass for each dynamical model based on 150 randomly selected walkers, while the shaded region represents the $1-\sigma$ uncertainty for the Fiducial model. The symbols show the different NGC~5128 literature enclosed mass measurements \citep{Hui1995,Mathieu1996,Kraft2003,Peng2004,Woodley2007,Woodley2010,Hughes2022}, as well as the lower limit from \citet{Pearson2022}. We have colored the symbols according to the kinematic tracer used to inferred the total enclosed mass, with the colors representing the same tracer as in Figure~\ref{fig:Tumpet plot}. \textit{Right:} Dark matter fraction ($f_{DM}$) comparison for all the different dynamical models of this work. Lines are colored the same as in the left panel. The vertical lines show the virial radius for each dynamical model.  }
    \label{fig:Mass_comparison_all_models}
\end{figure*}

Many previous estimates exist of the enclosed gravitational mass of NGC~5128 at a range of radii.  These estimates are summarized in  the left panel of Figure~\ref{fig:Mass_comparison_all_models} and compared to our new dynamical modeling.  

We find good agreement between the results of our Fiducial model and previous mass estimates at small galactocentric radius \citep{Hui1995,Kraft2003}. At larger radii, we find that our models have systematically lower masses than previous mass estimates made using GCs and dwarfs as tracers \citep{Woodley2007,Karachentsev2007,Woodley2010,Hughes2022} as well as the mass estimate from \citet{Pearson2022} based on modeling of the Dw3 stream. The comparison to PNe masses is more mixed, with a favorable match to the estimate from  \citet{Peng2004}, but higher masses than other previous estimates made using PNe as tracers \citep{Mathieu1996,Woodley2007}. These comparisons are consistent with the interpretation that our best-fit 
Fiducial model represents a statistical compromise among these tracers.

\subsection{Comparison to previous dark matter profile fits}
\label{subsec:comparison_to_lit_dark_matter_estimates}
Two previous works, \citet{Peng2004} and \citet{Pearson2022}, have estimated parameters for a standard NFW profile for NGC~5128. \citet{Peng2004} fitted PNe using an isotropic spherically symmetric Jeans model, obtaining an NFW scale density of $\log(\rho_{s})=-2.0$ and scale radius of $r_{s}=14.4$ kpc (no uncertainties reported). There is no mention of the ${\rm M}_{200}$ for the NFW profile in \citet{Peng2004}, only $c_{200}=12$. However, we can estimate an ${\rm M}_{200} = 6.32\times10^{11}$ M$_{\odot}$ based on their $r_{s}$ and $\rho_{s}$. These values are all inconsistent with those from our Fiducial model, producing a more centrally concentrated and much lower mass halo. For comparison to these results we run an isotropic ($\beta_{PNe}=0.0$) MCMC with only the PNe as kinematic tracers, but otherwise with the same assumptions as in our Fiducial model. For this isotropic PNe-only model, we find a scale density $\log(\rho_{s})=-2.01^{+0.23}_{-0.32}$ and scale radius of $r_{s}=14.9^{+4.1}_{-2.4}$ kpc, in very good agreement with the values found by \cite{Peng2004}. This comparison illustrates both the value of large-radius tracers for estimating the total halo mass, and, to a lesser degree, the effect of anisotropy on the fits.

\citet{Pearson2022} used the stellar stream of the dwarf satellite galaxy Dw3 (at a projected radius of $\sim 80$ kpc) to estimate a lower limit of the dark matter halo mass of NGC~5128 of ${\rm M}_{200} > 4.7\times10^{12}$ M$_{\odot}$, with an assumed $c_{200} = 8.6$ and stellar mass-to-light ratio of $\Upsilon_{K} = 0.7$. These values correspond to a more centrally concentrated, massive dark matter halo than our Fiducial model. The dynamical modeling of Dw3 done by \citet{Pearson2022} was tuned to reproduce the enclosed mass at 40 kpc found by \citet{Woodley2010} based on GC kinematics.

Of the models fit in the previous section, the one which is most comparable to this is the anisotropic ``GC only'' model (Section \ref{subsec:examining_uncertainties_data}). For this model, we find ${\rm M}_{200} = 5.0^{+5.8}_{-2.1} \times 10^{12}$ M$_{\odot}$, broadly consistent with the ${\rm M}_{200}$ lower limit from \citet{Pearson2022}. They conclude that the  morphology and central radial velocity for a single stream is insufficient to precisely determine halo parameters; revisiting this stream fitting with the updated tracer constraints now available would be worthwhile.

\section{Discussion}
\label{sec:discussion}
\subsection{Dark Matter Fractions}\label{sec:dmfrac}
Our dynamical model fitting in Section~\ref{sec:Model_results} show the need for large amounts of dark matter to explain the dynamics of the GCs, PNe, and dwarf satellite galaxies in NGC~5128, with a best-fit virial halo mass of ${\rm M}_{vir} = 4.4^{+2.4}_{-1.4} \times  10^{12} $M$_{\odot}$ and a stellar mass of $\sim 0.2 \times  10^{12} $M$_{\odot}$. Hence globally, the stellar mass fraction of the galaxy is found to be $\sim 4\%$ (formally $4.3^{+1.7}_{-1.3}\%$).

The stellar mass--halo mass relation has been measured for galaxies in a variety of ways, including abundance matching \citep[e.g.,][]{Moster2010}, weak lensing \citep[e.g.,][]{Leauthaud2012}, and from clustering analysis \citep[e.g.,][]{Zheng2007}.  A recent compilation of stellar mass--halo mass relations from \citet{Behroozi2019} shows that at NGC~5128's measured virial mass, the expected stellar mass fraction is 1--3\%. The higher end of that range is consistent with the measured stellar mass fraction within its uncertainty, while the lower end is less consistent, suggesting the possibility that NGC~5128 was more efficient than the average galaxy of this virial mass at transforming baryons into stars. Going the other direction, based on NGC~5128's stellar mass, the halo mass is expected to be $\gtrsim$10$^{13}$~M$_\odot$ based on these stellar mass--halo mass relations, which is clearly inconsistent with the measured virial mass.

Since observational measurements encompassing a large fraction of the virial radius are challenging, many observational studies have focused instead on quantifying the fraction of dark matter in the inner portions of galaxies, often measured at notional comparison radii of 1 and $5 R_{eff}$.  For the dark matter fraction ($f_{DM}$) in NGC~5128 within $1R_{eff}$ we find $f_{DM} = 0.11\pm0.04$ for our Fiducial model.  Previous dynamical modeling results for other early-type galaxies \citep{Atlas3Dcappellari}, along with spiral and S0 galaxies \citep{Williams2009}, show this to be a typical value, with most galaxies in these samples having $f_{DM} < 0.2$ at $1 R_{eff}$. Individual galaxy measurements for the giant elliptical
M87 and NGC~5846, more comparable to the approach we use for NGC~5128, also find $f_{DM} \sim 0.1$ within $1R_{eff}$ \citep{Zhu2014,Zhu2016}.  However, theoretical models \citep{Lovell2018} and other dynamical and lensing results \citep{Barnabe2011,Tortora2012} suggest higher dark matter fractions within $1R_{eff}$. This is a tension previously recognized. It could be due to inconsistencies in how sizes are measured for real galaxies compared to those in simulations, uncertainty about the initial mass function, or perhaps to true differences between the inner dark matter distribution in observed galaxies compared to simulations (see additional discussion in \citealt{Lovell2018}).
 
At larger radii, within $5R_{eff}$,  we find $f_{DM} = 0.52\pm0.08$. Previous measurements using tracer-mass estimator methods on large GC samples from the SLUGGS project  \citep{Alabi17} find a wide range of dark matter fractions within $5R_{eff}$ from  $f_{DM}$ of 0.1 to 0.95, with most higher than NGC~5128. Simulations generally find less scatter in $f_{DM}$ at these larger radii among galaxies than do observations, but do not agree on the actual value of $f_{DM}$. For example, \citet{Wu2014} find $\sim0.5$ for $f_{DM}$ for galaxies in the stellar mass range of NGC~5128, while more recent results from \citet{Remus2017} and \citet{Lovell2018} suggest much higher $f_{DM}$ $\gtrsim$0.7 with minimal scatter. We note that the methodology of existing observational measurements varies significantly, with many studies using basic spherical isotropic dynamical models with limited radial tracer coverage, and only a handful of previous estimates are based on Jeans modeling out to large radii as we present here.

\subsection{Concentration--Mass Relation}

Numerical simulations show that the structural properties of dark matter halos change as a function of mass, with higher mass halos becoming less concentrated, as halo density is correlated with the density of the universe at the time of formation \citep[e.g.][]{Navarro1997,Bullock2001}.  The relation between halo mass and concentration is thus an important tracer of the formation history of galaxy halos.  

As a prior on the NFW dark matter component of our NGC~5128 mass model, we adopted the concentration--mass ($c_{vir}-{\rm M}_{vir}$) relation of \citet{Bullock2001}.  This prior of $c_{vir}$=14.0$\pm$5.6 was estimated based on a previous NGC~5128 mass measurement \citep{Woodley2007}. 
The actual estimated value of $c_{vir}$ for our Fiducial model is $c_{vir}$=5.6$^{+2.4}_{-1.6}$ with a ${\rm M}_{vir}=4.4^{+2.4}_{-1.4}\times$10$^{12} {\rm M}_{\odot}$. That our best-fit concentration is in the tail of the prior indicates the data quite strongly favor the measured low concentration. 

As it happens, the measured value is in somewhat better agreement with the more recent concentration--mass relations of \citet{Munoz2011} and \citet{Dutton2014}, who find mean values of  $c_{vir}\sim8.3$ and $c_{vir}\sim9.2$ at our virial mass, respectively. While our Fiducial concentration value is still nominally lower than these predictions, the simulations exhibit significant halo to halo scatter in $c_{vir}-{\rm M}_{vir}$, of order $\sim$30\%. Hence the overall conclusion is that NGC~5128's inferred dark matter halo is consistent with the distribution of halo properties at this virial mass observed in cosmological simulations. A much larger set of measured halo concentrations for massive ellipticals would be necessary to more carefully compare simulations to observations.

\subsection{Halo mass--$N_{GC}$ Relation}

A number of papers have demonstrated a relationship between the number of GCs in a galaxy ($N_{GC}$) and its total (including dark matter) mass, which is apparently linear over a large range in halo masses (e.g., \citealt{Harris2017,Burkert2020}). It is notable that the claimed scatter in this relation is small ($< 0.3$ dex), which if true would make the counting of GCs one of the most reliable methods to estimate halo masses.

Since NGC~5128 has both a good estimate of its total GC population ($N_{GC} = 1450\pm160$ ; \citealt{Hughes2021}) and total mass from the present work (${\rm M}_{vir,t} = 4.7^{+1.4}_{-1.4} \times 10^{12} {\rm M}_{\odot}$), it presents a compelling test of this relation.
We use the two most recent derived versions of the relation, from \citet{Harris2017} and \citet{Burkert2020}. For the former, the predicted mass from the measured $N_{GC}$ is $\sim 1.6 \times 10^{13} {\rm M}_{\odot}$, with an uncertainty of about a factor of 2 from the stated scatter in the relation. The offset of NGC~5128 from the relation is about 0.54 dex, or nearly $2\sigma$ on the low end. For the \citet{Burkert2020} relation, the inferred halo mass is lower, about $7 \times 10^{12} {\rm M}_{\odot}$, which is higher than but in less tension than our NGC~5128 mass measurement.  

As we discussed above for the measurement of halo concentration, it is impossible to reach a firm conclusion about variance in one of these relations from a single measurement. Nonetheless, given the small number of galaxies for which there are robust measurements of both their virial masses and GC systems, revisiting the scatter in the $N_{GC}$--total mass relation should be a priority once more virial mass estimates are available.

\subsection{The Mass of NGC~5128 In Context}

NGC~5128 is one of the more massive galaxies out to a distance of $\sim 5$ Mpc, making it an important benchmark system for connecting its stellar and dark matter halo properties to expectations from the $\Lambda$CDM model and other prominent, nearby galaxies.  

Within the Local Group, the masses of the Milky Way and M31 have been the subject of many studies. %, but have been relatively difficult to pin down with different techniques and data sets yielding answers that differ significantly.  
For the Milky Way, astrometric data from the Gaia mission has provided critical tangential velocities for many mass tracers (satellites galaxies, GCs and halo stars; see e.g. \citealt{Callingham19,Eadie19,Watkins19,Li20,Deason21}), improving agreement between recent studies, most of which point to total masses around 1--1.5$\times$10$^{12}$ M$_{\odot}$ \citep[see][for a recent review]{WangMW}.  This mass of the Milky Way is similar to mass estimates of other nearby, massive spiral galaxies; M$_{vir,\rm M83}$$\approx$0.9$\times$10$^{12}$ M$_{\odot}$ \citep{K07}; M$_{vir, \rm M81}$$\approx$1$\times$10$^{12}$ M$_{\odot}$ \citep{K06}. This places NGC~5128 at about three to four times more massive than the Milky Way and other nearby spirals.

In M31, modern measurements have yielded virial mass values in the range ${\rm M}_{vir,\rm M31}$=0.5--3.5$\times$10$^{12}$ M$_{\odot}$, including satellite galaxy \citep{Watkins10,Hayashi14}, globular cluster \citep[e.g.][]{Veljanoski13} and stellar stream \citep{Fardal13,Dey23} kinematic measurements; see the recent compilation of \citet{Patel23}.
%; see figure~4 in \citet{Patel23} for a recent compilation. 
In that same work, \citet{Patel23} combine proper motion measurements of four M31 satellites and cosmological simulations to infer ${\rm M}_{vir,M31}$=2.85$\times$10$^{12}$ M$_{\odot}$.   This value is within a factor of $\sim$1.5 of the NGC~5128 virial mass, and the uncertainties of the two measurements overlap. Those authors also note that M31 is on the upper end of expectations for the stellar mass-halo mass relation, as we have seen for NGC~5128 in Section~\ref{sec:dmfrac}.  Future measurements are necessary to confirm this most recent, relatively high mass for M31.  Nonetheless, it is plausible that M31 and NGC~5128 have similar masses.

Kinematic measurements out to large radii ($\gtrsim$50--100 kpc) using GCs, PNe and satellite galaxies is feasible for more galaxies in the nearby universe.  A systematic data set that pursued this goal would be valuable for constraining the virial masses  and mass profiles of the nearest massive galaxy halos in a consistent way, allowing for further intercomparisons and comparisons with $\Lambda$CDM.

\section{Summary}
We used the discrete Jeans anistoropic model CJAM \citep{Watkins2013} to estimate the halo parameters of NGC~5128 using its GCs, PNe, and dwarf satellite galaxies as kinematic tracers out to a radius of 250~kpc, with the tracer anisotropies and stellar mass-to-light ratio fit as free parameters.

Our Fiducial model fit has a dark matter halo virial mass of ${\rm M}_{vir}=4.4^{+2.4}_{-1.4}\times10^{12} {\rm M}_{\odot}$, a concentration of $c_{vir}=5.6^{+2.4}_{-1.6}$, and a corresponding virial radius of $r_{vir}=432^{+66}_{-51}$ kpc. This halo constitutes only $\sim10\%$ of the total mass at $1R_{eff}$ but $\sim51\%$ at $5R_{eff}$. This model fits nearly all the data, with the exception of a subset of the PNe that are potentially not in virial equilibrium, and the derived halo parameters are relatively insensitive to modest variations in the tracer anisotropies and system inclination.

NGC~5128 appears to sit ``low" on the mean stellar mass--halo mass and GC mass--halo mass relations, which both predict a halo virial mass closer to ${\rm M}_{vir} \sim 10^{13} {\rm M}_{\odot}$. No general statements about these relations can be made with a single galaxy, but our analysis highlights the value of comprehensive dynamical modeling of nearby galaxies, and the importance of using multiple tracers to allow cross-checks for model robustness.

\section*{Acknowledgement}

Work by A.D.~and A.C.S.~has been supported by NSF AST-1813609.  AKH and DJS acknowledge support from NSF grants
AST-1821967 and 1813708.  NC acknowledge support by NSF grant AST-1812461. JS acknowledges support from NSF grants AST-1514763 and AST-1812856 and the Packard Foundation. Support for SP work was provided by NASA through the NASA Hubble Fellowship grant \#HST-HF2-51466.001-A awarded by the Space Telescope Science Institute, which is operated by the Association of Universities for Research in Astronomy, Incorporated, under NASA contract NAS5-26555.  

This paper made use of the the MGE fitting method and software by \citet{MultiGaussian_Cappellari} and JAM modelling method of \citet{JAM_cappellari}, as well as the \rm astropy package \citep{Astropy2013}.

\newpage
\begin{deluxetable}{clll}
\caption{MGE Parametrization of NGC~5128}
\label{tab:MGEs}
\fontsize{6}{6}\selectfont
\tablecolumns{4}
\tablehead{Component & \colhead{Projected Lum.\tablenotemark{a}} & \colhead{$\sigma$\tablenotemark{b}} & \colhead{Axis ratio\tablenotemark{c}} \\
 & \colhead{($L_{\odot}/{\rm pc}^{2}$)} & (arcsec) &  }
\startdata
 1......& 166545.36 & 1.008 & 0.967 \\
 2......& 60326.79 & 3.685 & 0.999 \\
 3......& 19415.50 & 11.672 & 0.935 \\
 4......& 5708.00 & 31.180 & 0.868 \\
 5......& 2142.51 & 41.267 & 0.800 \\
 6......& 1387.80 & 78.027 & 0.800 \\
 7......& 612.72 & 120.771 & 0.800 \\
 8......& 248.34 & 190.376 & 0.800 \\
 9......& 174.35 & 273.093 & 0.800 \\
 10.....& 20.27 & 431.253 & 0.802 \\
 11.....& 48.95 & 520.122 & 0.800 \\
 12.....& 12.34 & 935.011 & 0.800 \\
 13.....& 2.14 & 1615.597 & 0.800 \\
 14.....& 0.28 & 2493.896 & 0.800 \\
 15.....& 0.12 & 3682.893 & 0.800 \\
\enddata
\tablecomments{List of the fifteen components of the MGE model parametrization of the surface brightness of NGC~5128. The MGE model assumes PA $=35^{\circ}$. The total K-band luminosity for the MGE model is $L_{K} = 1.544\times10^{11} L_{\odot}$.}
\tablenotetext{a}{Central surface brightness of component.}
\tablenotetext{b}{Gaussian $\sigma$ along major axis of component.}
\tablenotetext{c}{Fitted semi-major axis ratio $b'/a'$ of component.}
\end{deluxetable}

%%%% TABLE FOR Fiducial BEST FIT PARAMETERS
\begin{rotatetable}
%\movetableright=5mm
\begin{deluxetable}{lccccccc}
\tablecaption{Mass Modeling Results}
\label{tab:Best_fit_values}
%\fontsize{6}{6}\selectfont
\tablecolumns{8}
\tablehead{\colhead{Model} & \colhead{$\Upsilon_{K}$\tablenotemark{a} }& \colhead{$\log(\rho_{s})$\tablenotemark{b} }& \colhead{$r_{s}$ \tablenotemark{c}} & \colhead{$\beta_{red}$ \tablenotemark{d}} & \colhead{$\beta_{blue}$ \tablenotemark{e}} & \colhead{$\beta_{PNe}$\tablenotemark{f}} & \colhead{$\beta_{dwarf}$\tablenotemark{g}} \\ & \colhead{($10^{12}{\rm M}_{\odot}$)} & \colhead{$\log({\rm M}_{\odot}{\rm pc}^{-3})$} & \colhead{(kpc)} & \colhead{} & \colhead{} &  \colhead{}& \colhead{}}
\startdata
       Fiducial & $1.6^{+0.2}_{-0.3}$ &  $-3.14^{+0.37}_{-0.33}$ & $77.5^{+44.5}_{-28.9}$ &  $-0.28^{+0.08}_{-0.07}$ & $-0.34^{+0.07}_{-0.05}$ & $-0.11^{+0.08}_{-0.08}$ & $-0.13^{+0.27}_{-0.18}$ \\
       \hline
       GCs only & $1.3^{+0.3}_{-0.3}$ & $-3.02^{+0.41}_{-0.52}$ & $76.7^{+88.7}_{-33.1}$ & $-0.26^{+0.09}_{-0.08}$ & $-0.32^{+0.08}_{-0.06}$ & \nodata & \nodata \\
       PNe only & $1.3^{+0.2}_{-0.2}$ & $-1.91^{+0.23}_{-0.26}$ & $12.8^{+4.7}_{-3.1}$ & \nodata & \nodata & $-0.20^{+0.08}_{-0.09}$ & \nodata \\
       Dwarfs only & $1.1^{+0.3}_{-0.3}$ & $-2.45^{+0.45}_{-0.74}$ & $36.9^{+40.5}_{-12.9}$ & \nodata & \nodata & \nodata & $-0.09^{+0.28}_{-0.23}$ \\
       Isotropic &  $1.2^{+0.3}_{-0.3}$ & $-2.61^{+0.27}_{-0.28}$ & $43.6^{+19.3}_{-12.3}$ & \nodata & \nodata & \nodata & \nodata \\
       Inclination = $60^{\circ}$ &  $1.5^{+0.3}_{-0.3}$ & $-2.91^{+0.32}_{-0.35}$ & $58.7^{+35.2}_{-19.1}$ & $-0.30^{+0.09}_{-0.07}$ & $-0.33^{+0.07}_{-0.05}$ & $-0.10^{+0.09}_{-0.11}$ & $-0.06^{+0.27}_{-0.24}$ \\    
       Cored &  $2.3^{+0.1}_{-0.1}$ & $-2.87^{+0.22}_{-0.22}$ & $131.4^{+32.1}_{-23.1}$ & $-0.32^{+0.07}_{-0.06}$ & $-0.35^{+0.06}_{-0.04}$ & $-0.20^{+0.08}_{-0.09}$ & $-0.15^{+0.26}_{-0.17}$ \\  
\enddata
\vspace{2mm}
\tablecomments{The best-fit parameters for all the dynamical models. The Fiducial model is first and highlighted in black (Section \ref{subsec:Fiducial_results}). The next set of fits are for variations in the tracer population (Section \ref{subsec:examining_uncertainties_data}), and the final set of fits are variations in the model assumptions (Section \ref{subsec:examining_uncertainties}).}

\tablenotetext{a}{The $K$-band stellar mass-to-light ratio $\Upsilon_{K}$.}
\tablenotetext{b}{The $\log$ NFW scale density.}
\tablenotetext{c}{The NFW scale radius $r_{s}$.}
\tablenotetext{d}{Velocity anisotropy $\beta$ for the red GCs.}
\tablenotetext{e}{Velocity anisotropy $\beta$ for the blue GCs.}
\tablenotetext{f}{Velocity anisotropy $\beta$ for the PNe.}
\tablenotetext{g}{Velocity anisotropy $\beta$ for the dwarf satellite galaxies.}
\end{deluxetable}
\end{rotatetable}

%%%% TABLE FOR MODEL DERIVED QUNATITIES

\begin{deluxetable}{lllllll}
\tablecaption{Mass Modeling Derived Quantities}
\label{tab:derived_quantities}
\fontsize{6}{6}\selectfont
\tablecolumns{7}
\tablehead{Model & ${\rm M}_{vir}$\tablenotemark{a} & $c_{vir}$\tablenotemark{b} & ${\rm M}_{\ast}$\tablenotemark{c} & ${\rm M}_{vir,t}$\tablenotemark{d} & $f_{DM,1 R_{eff}}$\tablenotemark{e} & $f_{DM,5 R_{eff}}$\tablenotemark{f} \\ & ($10^{12}{\rm M}_{\odot}$) &  & ($10^{11}{\rm M}_{\odot}$) & $10^{12}{\rm M}_{\odot}$ & $(\%)$ & $(\%)$}
\startdata
       Fiducial & $4.4^{+2.4}_{-1.4}$ & $5.6^{+2.4}_{-1.6}$ & $1.9^{+0.2}_{-0.3}$ & $4.7^{+1.4}_{-1.4}$ & $10.7^{+3.9}_{-3.6}$  & $51.9^{7.5}_{-8.4}$ \\
       \hline
       GCs only & $6.1^{+7.3}_{-2.6}$ & $6.3^{+3.0}_{-2.6}$ & $1.7^{+0.3}_{-0.3}$ & $6.1^{+3.9}_{-1.9} $ & $14.0^{+5.6}_{-4.5}$ & $60.6^{+7.1}_{-7.6}$ \\
       PNe only & $0.7^{+0.3}_{-0.2}$ & $17.7^{+4.2}_{-3.8}$ & $1.7^{+0.2}_{-0.2}$ & $0.8^{+0.2}_{-0.2} $ & $19.3^{+4.5}_{-4.6}$ & $51.3^{+7.1}_{-7.8}$ \\
       Dwarfs only & $3.3^{+2.0}_{-1.2}$ & $10.8^{+5.5}_{-5.5}$ & $1.4^{+0.3}_{-0.3}$ & $3.5^{+ 1.6}_{-1.0}$ & $22.1^{+10.8}_{- 9.8}$ & $69.2^{+7.5}_{-13.2}$ \\
       \hline
       Isotropic & $3.7^{+1.4}_{-0.9}$ & $9.3^{+2.7}_{-2.2}$ & $1.6^{+0.3}_{-0.3}$ & $3.8^{+0.8}_{-0.8}$ & $18.9^{+6.5}_{-4.5}$ & $66.3^{+5.9}_{-7.6}$\\
       Inclination = $60^{\circ}$ & $3.7^{+1.9}_{-1.0}$ & $7.0^{+2.5}_{-2.0}$ & $1.8^{+0.3}_{-0.3}$ & $3.9^{+1.4}_{-0.9}$ & $12.5^{+4.6}_{-3.5}$ & $56.2^{+6.1}_{-7.1}$ \\    
       Cored & $4.5^{+2.1}_{-1.4}$ & $3.3^{+0.4}_{-0.4}$ & $2.6^{+0.1}_{-0.1}$ & $4.8^{+1.8}_{-1.2}$ & $0.5^{+0.3}_{-0.2}$ & $23.2^{+6.6}_{-4.9}$\\  
\enddata
\vspace{2mm}
\tablecomments{The best-fit parameters for all the dynamical models. The Fiducial model is first and highlighted in black (Section \ref{subsec:Fiducial_results}). The next set of fits are for variations in the tracer population (Section \ref{subsec:examining_uncertainties_data}), and the final set of fits are variations in the model assumptions (Section \ref{subsec:examining_uncertainties}).}

\tablenotetext{a}{The virial mass ${\rm M}_{vir}$ .}
\tablenotetext{b}{The halo concentration $c_{vir} = r_{vir}/r_s$.}
\tablenotetext{c}{The stellar mass of the galaxy ${\rm M}_{\ast}$.}
\tablenotetext{d}{The total mass ${\rm M}_{vir,t}$ (${\rm M}_{vir} + {\rm M}_{\ast})$.}
\tablenotetext{e}{The dark matter fraction within $1 R_{eff}$.}
\tablenotetext{f}{The dark matter fraction within $5 R_{eff}$.}
\end{deluxetable}

\newpage
\appendix
\setcounter{figure}{0}  
\renewcommand\thefigure{A\arabic{figure}} 
\label{appendix}
${\rm M}_{vir}$ is the mass enclosed at a radius where the mean halo density ($\frac{3}{4\pi R^{3}_{vir}} \int^{r_{vir}}_{0} \rho_{r}4\pi r^{2}dr$) is $\delta_{th}\times\Omega_{0}$ times the critical density ($\rho_{cri}$) of the universe at that redshift. $\delta_{th}$ is the oversize density of a collapse object in the ``top-hat'' collapse model, and $\Omega_{0}$ the matter contribution to the critical density. 
To calculate $\rho_{cri}$, $\delta_{th}$ and $\Omega_{0}$ we use the {\em Astropy.cosmology} python class with the cosmology based on the Planck mission \citep{PlanckCollaboration2015}. At the redshift of NGC~5128 (z$=0.00183$\footnote{from the NASA/IPAC Extragalactic Database (NED)}) we get a value of $\rho_{cri}\approx1.27\times10^{-7}$ M$_{\odot}$/pc$^{3}$, $\delta_{th}\approx331.50$, and $\Omega_{0}\approx 0.308$.
Finally, M$_{vir}$ is given by:
\begin{equation}
    {\rm M}_{vir} = \frac{4\pi}{3}\delta_{th}\Omega_{0}\rho_{cri}R^{3}_{vir}
\end{equation}
We emphasise the difference between ${\rm M}_{vir}$, $c_{vir}$, and $r_{vir}$ with the also commonly use quantities ${\rm M}_{200}$, $C_{200}$, and $R_{200}$ calculated for a mass enclosed at the distance $200\times\rho_{cri}$ regardless of the redshift. The virial quantities for this redshift are typically smaller, with ${\rm M}_{200} \sim 0.8 \times {\rm M}_{vir}$.

\end{document}